%% file: apaper.tex
\documentclass[prb,aps,twocolumn,reprint,superscriptaddress,titlepage,floatfix,showkeys]{revtex4-1}
\usepackage{graphicx}
\usepackage{nicefrac}
\usepackage{amsfonts}
\usepackage[version=3]{mhchem}
\usepackage{amssymb}
\usepackage{amsmath} 
\usepackage{subfigure}
\usepackage{multirow} 
\usepackage{tabularx} 
\usepackage{array}
\usepackage{units}
\usepackage{braket}
\usepackage{bm,times}
\usepackage{booktabs}
\usepackage{enumitem}
\usepackage{mathtools}
\usepackage{epstopdf}
\usepackage{threeparttable}
\usepackage[colorlinks = true, linkcolor = blue, urlcolor  = blue, citecolor = blue, anchorcolor = blue]{hyperref}

\def\be{\begin{equation}}
\def\ee{\end{equation}}
\def\bea{\begin{eqnarray}}
\def\eea{\end{eqnarray}}

\def\dg{$^\circ$}

\def\asi{{\it a}-Si}
\def\age{{\it a}-Ge}

\def\asih{{\it a}-Si:H}

\begin{document}

\title{Disorder by design: A data-driven approach to amorphous 
semiconductors without total-energy functionals} 

\author{Dil K. Limbu}
\email{dil.limbu@usm.edu}
\affiliation{Department of Physics and Astronomy, The University of 
Southern Mississippi, Hattiesburg, Mississippi 39406, USA}

\author{Stephen R. Elliott} 
\email{sre1@cam.ac.uk}
\affiliation{Department of Chemistry, University of Cambridge, 
Cambridge CB2 1EW, United Kingdom}

\author{Raymond Atta-Fynn}
\email{r.attafynn@gmail.com}
\affiliation{Department of Physics, The University of Texas at Arlington, Texas 76019, USA}

\author{Parthapratim Biswas}
\email[Corresponding author:]{partha.biswas@usm.edu}
\affiliation{Department of Physics and Astronomy, The University of Southern 
Mississippi, Hattiesburg, Mississippi 39406, USA}

\begin{abstract}
This paper addresses a difficult inverse problem that involves 
the reconstruction of a three-dimensional model of tetrahedral 
amorphous semiconductors via inversion of diffraction data. By 
posing the material-structure determination as a multi-objective 
optimization program, it has been shown that the problem can be 
solved accurately using a few structural 
constraints, but no total-energy functionals/forces, which 
describe the local chemistry of amorphous networks. The approach 
yields highly realistic models of {\asi}, with no or only a few 
coordination defects ($\le$ 1\%), a narrow bond-angle 
distribution of width 9--11.5{\dg}, and an electronic gap 
of 0.8--1.4 eV. These data-driven information-based models 
have been found to produce electronic and vibrational properties 
of {\asi} that match accurately with experimental data and rival 
that of the Wooten-Winer-Weaire (W3) models. 
The study confirms the effectiveness of a multi-objective optimization 
approach to the structural determination of complex materials, 
and resolves a long-standing dispute concerning the uniqueness 
of a model of tetrahedral amorphous semiconductors obtained 
via inversion of diffraction data. 
\end{abstract}

\maketitle 

\section*{Introduction}
Experimental information from scattering measurements, such as X-ray, 
electron and neutron diffraction, play an important role in the 
structural determination of ordered and disordered 
materials.~\cite{Guinier1955,Warren1969} 
Diffraction experiments typically provide scattering intensities 
from the constituent atoms in the wavevector space, which are related 
to the two-body correlation function between atoms in real space via 
the Fourier transform of the structure factor. For amorphous solids, 
the reconstruction of the correct three-dimensional real-space 
structure from scattering data is an archetypal example of inverse problems in 
computational modeling of materials. 
Since the lack of information 
from higher-order atomic-correlation functions cannot be remedied 
by any amount of computational/mathematical trickeries, it is 
absolutely necessary to supplement experimental data by additional 
information to determine a unique structural model of a material. 
Among earlier approaches to address the problem, the reverse 
Monte Carlo (RMC) method~\cite{McGreevy1988} is an elegant
method that relies on the generation of Markov 
chains or random walks to optimize a suitable cost function 
in the state-vector or configurational space.  
While a variant of the RMC method was used 
by Strong and Kaplow~\cite{Strong1968,Kaplow1968} as early as the 
1960s to predict the structure of crystalline B$_2$O$_3$~\cite{Strong1968} 
and vitreous selenium,~\cite{Kaplow1968} by refining 
X-ray diffraction data via random walks, it was McGreevy and 
co-workers~\cite{McGreevy1988, Keen1990, McGreevy2001} who first 
applied the method systematically to model the structure 
of disordered solids.~\cite{McGreevy1988}  Since then, 
the method has been widely used to study a variety 
of complex disordered systems, including liquids, glasses, 
disordered alloys, and proteins.~\cite{McGreevy2001} 

Despite significant efforts over the past decades, 
direct application of the RMC method to determine reliable structural 
solutions of disordered materials from their diffraction data 
has remained a challenging problem to date. The problem is particularly 
acute for amorphous tetrahedral semiconductors, such as 
amorphous silicon ({\asi}) and germanium ({\age}).  
Although a number of RMC or RMC-derived studies~\cite{McGreevy1988, Keen1990, Gereben1994, McGreevy2001, Walters1996, Biswas2004, 
Tucker2007, Opletal2013, Cliffe2010, Cliffe2017} on {\asi}/{\age} 
have been reported in the past, none of these could demonstrate 
the presence of a gap in the electronic spectrum 
and a low concentration of 
coordination defects ($\le$ 1\%), as observed in optical 
measurements~\cite{Klazes1982,Kageyama2011} and electron 
spin resonance (ESR)~\cite{Brodsky1969} experiments, respectively, for 
these materials.  Thus, the problem of the structural determination 
of tetrahedral amorphous semiconductors from diffraction 
data {\em without} using a total-energy functional
remains unsolved up until now. 
Since an ensemble of 
three-dimensional structures can lead to an identical 
two-body correlation function, additional 
information is required to uniquely determine the correct 
structure of a material by suitably reducing the volume
of the solution space.  
While this reduction is generally achieved by imposing 
structural constraints during RMC simulations, the 
hierarchy and conflictive nature of the constraints render 
the resulting optimization problem very difficult, leading 
to poor structural solutions. To overcome this problem, 
a new breed of hybrid approaches have been developed in 
recent years~\cite{Biswas2005,Biswas2007,Pandey2015,Pandey2016b,Limbu2018}, 
which can successfully address the uniqueness problem by 
simultaneously employing experimental data and a total-energy 
functional. However, these hybrid approaches crucially rely 
on the availability of suitable total-energy functionals 
and often, for many multinary systems, it is difficult 
to guide the hybrid solutions without resort to the use of 
expensive quantum-mechanical force fields.  
This considerably 
increases the computational complexity of the problem, 
which limits the applicability of {\it ab initio}-based 
hybrid approaches in addressing large disordered systems. 
This necessitates the development of purely data-driven 
information-based approaches to material simulations, 
which rely on experimental data and auxiliary structural 
information. 

The principal aim of this paper is to demonstrate conclusively 
that an accurate structural solution for tetrahedral amorphous 
semiconductors can be obtained {\em without} the need for a 
total-energy functional or atomic forces but using diffraction 
data only, assisted by a few structural constraints. 
By developing an efficient constraint optimization scheme, within 
the framework of Monte Carlo methods, which scales linearly 
with the system size, we show that 
tetrahedral models of {\asi} can be constructed
by the judicious use of local structural/chemical constraints 
and diffraction data. More importantly, we demonstrate 
that the data-driven information-based models are 
energetically stable in the sense that the models 
correspond to a stable local minimum of a quantum-mechanical 
total-energy functional and that the models produce 
a clean gap in the electronic spectrum. 
Furthermore, we show that the resulting models exhibit structural, electronic, 
and vibrational properties of {\asi} that match excellently with 
the corresponding experimental data from X-ray, Raman spectroscopy, 
and inelastic neutron-diffraction measurements, as well as 
those obtained from using conventional simulation techniques, 
based on classical and quantum-mechanical force fields. 
Finally, we discuss an extension of the scheme with 
a few examples to incorporate microstructural properties of 
realistic samples of {\asi} and {\age} as observed in experiments.

\section*{Results and Discussion}
In discussing our results, we first demonstrate the superior structural 
quality of the new models over existing RMC or RMC-like models.  
This is followed by an extensive comparison of the results 
from our models with those from conventional Monte Carlo and 
molecular-dynamics simulations, using classical, quantum-mechanical, 
and machine-learning-based potentials. We have placed great 
emphasis on the electronic structure of {\asi}, as it is the electronic 
properties that are most difficult to produce accurately, not only 
in RMC/RMC-like methods but also in MD and MC simulations, and 
are often not discussed in the literature with sufficient detail. 
We then proceed to compare our results with hybrid models that employ 
both total-energy functionals and experimental data. 
This is followed by new ideas that serve as natural corollaries 
of the current approach in an effort to 
address microstructural properties of experimentally 
obtained {\asi} samples, which cannot be described using 
conventional continuous random network (CRN) models 
of {\asi}. Since the multi-objective 
constraint optimization 
problem in the present study was handled by using constraint 
Monte Carlo (CMC) methods, we shall refer to our models as 
CMC19 hereafter. 

\subsection{Comparison with earlier RMC-derived models} 
To examine the accuracy of structural quantities and the 
novelty of the CMC19 approach, we begin by listing the 
characteristic properties of the CMC19 model in Table 
\ref{TAB1}, along with those obtained from various 
RMC~\cite{Walters1996,Biswas2004} and 
RMC-derived~\cite{Cliffe2010,Gereben2011,Cliffe2017} approaches. 
A salient feature of all these methods is that none of 
the methods uses any total-energy functionals during 
structural formation but only experimental data in conjunction 
with a few constraints.  The results clearly indicate 
that the CMC19 models outperform all other models, as far 
as the values of $\Delta \theta$, $c_4$, and $E_g$ in 
Table \ref{TAB1} are concerned. 
\begin{table}[t!]
\caption{\label{TAB1}
Comparison of results from various information-based approaches. 
$N$, $\langle\theta\rangle$, $\Delta\theta$, $c_4$ and $E_g$ 
indicate the size of the system, the average bond angle, the 
RMS width of bond angles, the percentage of four-fold-coordinated 
atoms and the value of the electronic gap, respectively.
}
\begin{ruledtabular}
\begin{tabular}{lccrrc}
Model &N &$\langle\theta\rangle$ &$\Delta\theta$ & $c_4$ & $E_{g}$ (eV) \\
\hline
CMC19 &216 &109.08 &10.95 &100 & 1.18 \\
CMC19 &512 &109.14 &10.61 &99.22 & 1.09 \\
SOAP\footnote{Estimated values from supplementary information 
in Ref.\,\citenum{Cliffe2017}.\label{soap}} &512 &NA & NA & 95  & None\\
EX-INVERT\footnote{From Ref.\;\citenum{Gereben2011}.} &216 &NA & NA & 94-97  & NA\\
INVERT\footnote{From Ref.\,\citenum{Cliffe2010}} &512 &NA & NA & 92\textsuperscript{\ref{soap}}   & None\\
RMC04\footnote{From Ref.\;\citenum{Biswas2004}.} &500 &109.01 & 12.5 & 88  & None \\
RMC96\footnote{Results for {\age} from Ref.\;\citenum{Walters1996}.} &3000 &109.4 & 8.5 & 52 & None
\end{tabular}
\end{ruledtabular}
\end{table}
Although the lack of precise information 
on the bond-angle distribution 
from INVERT and SOAP models, in particular the root-mean-square 
(RMS) bond-angle width, $\Delta \theta$, makes it difficult 
to compare our results on a quantitative footing, it is evident 
that the INVERT and SOAP models produce too many defects 
(about 5--8\%) to open a gap in the respective electronic spectrum. 
It may be noted that the structural and electronic quality of 
a model {\asi} network is primarily determined by -- 
apart from the structure factor or its real space 
counterpart -- a trinity of three 
quantities: a) the width of the bond-angle distribution 
($\Delta \theta$); b) the concentration of four-fold-coordinated 
atoms ($c_4$); and c) the value of the electronic gap ($E_g$). 
It is therefore crucial for a realistic model of {\asi} to exhibit a 
small value of $\Delta\theta$ (of about 9--12{\dg}) 
and a large value of $c_4$ (typically $\ge$ 98\%). While an extension of the INVERT 
approach, EX-INVERT,~\cite{Gereben2011} is reported to 
have produced highly four-fold-coordinated networks with $c_4$
values of up to 97\%, the author provided no data on the width 
of the bond-angle distribution and the statistics of 
atomic coordination in Ref.\,\onlinecite{Gereben2011}. 
Given the multi-objective nature of the problem, it 
is possible to construct a network with a low 
concentration of coordination defects at the expense 
of a high value of $\Delta \theta$, and vice versa. 
However, such networks may not represent a stable 
physical solution upon total-energy relaxation. Thus, 
a structural model from a data-driven information-based 
approach must satisfy the aforementioned criteria 
simultaneously in order to be compliant with 
experimental data from Raman and optical measurements, 
and electron spin resonance experiments. By contrast, 
the CMC19 models obtained in the present study clearly 
and unambiguously satisfied all these criteria.  It is 
remarkable that the 216-atom CMC19 model produces a 
100\% four-fold-coordinated network with an RMS width 
of bond angles of about 10.95{\dg} and an electronic 
gap of magnitude 1.18 eV. We emphasize that this is 
the first ever data-driven information-based model that 
produces no defects and a clean electronic gap, without 
using a total-energy functional during structural 
formation. This constitutes one of the major outcomes of the present 
study. In the following section, we shall demonstrate 
that even the unrelaxed CMC19 models of {\asi} are capable 
of producing a gap in the electronic spectrum and that the 
structural properties of the models are energetically 
stable, i.e., almost independent of {\it ab initio} 
total-energy relaxations. 
 
\subsection{Comparison with total-energy-based models}

In order to demonstrate the efficacy of the data-driven 
CMC19 approach over conventional approaches, using 
total-energy and forces, we now compare the results 
with an array of models from molecular dynamics 
and Monte Carlo simulations. While there exist a 
plethora of such models in the literature, we confine 
ourselves to few high-quality {\it a}-Si models 
that can produce accurate structural, electronic, and 
vibrational properties of {\asi}. 
To this end, we chose four representative models for 
comparison: 1) a classical molecular-dynamics (MD) 
model~\cite{JCP2018} based on the modified Stillinger-Weber 
potential~\cite{Vink2001}; 2) an MD model based on 
machine-learning (ML) potential~\cite{Deringer2018}; 
c) an {\it ab initio} MD model~\cite{Stich1991}; and 
d) a model obtained from using the Wooten-Winer-Weaire 
(W3)~\cite{W3} algorithm, modified by Barkema and 
Mousseau (BMW3).~\cite{Barkema2000} The latter can 
produce large 100\% defect-free configurations of {\asi}, 
and it is often considered as the benchmark for 
high-quality {\asi} models from simulations. These 
four models will be collectively referred to as 
{\em total-energy-based} (TEB) models. 

Table \ref{TAB2} presents an overview of structural 
properties of the TEB models, along with the CMC19 
models of size from 216 to 1000 atoms  before and 
after {\it ab initio} relaxations. It is apparent 
that the structural quality of the CMC19 models 
is on a par with the BMW3 model, and that the former are 
considerably better than those derived from AIMD 
and MLMD models, as far as the four-fold coordination 
of the models are concerned.  This observation equally 
applies to the electronic quality of the models, which 
is reflected in the size of the electronic gap. 
Considering the long simulation time ($\ge 20$ ns) needed 
to produce the SWMD models and the complexity associated 
with generating a machine-learning potential for {\asi}, 
it is evident that the CMC19 approach produces {\asi} 
models par excellence, using diffraction data and 
constraints only. Not only do the CMC19 models show the 
presence of an essentially clean gap in the electronic 
spectrum, but also the size of the gap, $E_g$, matches 
closely with the value obtained from the BMW3 model.
A relatively small value of $E_g$ for the 
1000-atom CMC19 model can be attributed to the presence of 
0.9\% coordination defects and a somewhat larger value of 
$\Delta \theta$, compared to its BMW3 counterpart, which 
affect the band-edge states. 
\begin{table}[t!]
\caption{\label{TAB2}
Comparison of CMC19 models with the best available models 
(of {\asi}) in the literature. Symbols have the same meaning 
as in Table \ref{TAB1}. $E_g$ and bond angles are expressed 
in the unit of electron-volt (eV) and degree, respectively. 
}
\begin{ruledtabular}
\begin{tabular}{lcccccccc}
\multicolumn{9}{c}{{\em Unrelaxed} CMC19 models} \\ 
\hline
Model &N &${\langle\theta\rangle}$ &$\Delta\theta$ & $c_4$ & $c_2$ & $c_3$ &$c_5$ & $E_{g}$ \\
\hline
CMC19 & 216  & 109.12  & 10.68  & 99.07  & 0  & 0.93 & 0 & 1.14 \\
CMC19 & 300  & 109.14  & 10.56  & 99.33  & 0  & 0.67 & 0 & 1.06 \\
CMC19 & 512  & 109.19  & 10.70  & 98.44  & 0  & 1.56 & 0 & 0.88 \\
CMC19 & 1000 & 109.13  & 11.15  & 99.10  & 0.30  & 0.60 & 0 & 0.56 \\
\hline \hline 
\multicolumn{9}{c}{{\em Ab initio-relaxed} CMC19 models} \\ 
\hline
Model &N &${\langle\theta\rangle}$ &$\Delta\theta$ & $c_4$ & $c_2$ & $c_3$ &$c_5$ & $E_{g}$ \\
\hline
CMC19 & 216  & 109.08 & 10.95 & 100   & 0    & 0    &0    &1.18 \\
CMC19 & 300  & 109.16 & 10.19 & 99.33 & 0    & 0.67 &0    &1.39 \\
CMC19 & 512  & 109.14 & 10.61 & 99.22 & 0    & 0.78 &0    &1.09 \\
CMC19 & 1000 & 109.08 & 11.43 & 99.10 & 0.30 & 0.40 &0.20 &0.76\\
\hline \hline
\multicolumn{9}{c}{{\em Total-energy-based} models} \\
\hline
Model &N &${\langle\theta\rangle}$ &$\Delta\theta$ & $c_4$ & $c_2$ & $c_3$ &$c_5$ & $E_{g}$ \\
\hline
BMW3\footnote{From Ref.~\onlinecite{Barkema2000}} & 512 & 109.14 &10.36  & 100 & 0 & 0 & 0 & 1.32\\
SWMD & 512 &109.27 &9.12 &99.22 & 0 & 0.39 & 0.39 & 1.01\\
MLMD\footnote{From Ref.~\onlinecite{Deringer2018}} & 512 &109.19 &9.69 &98.44 & 0 & 0.78 &0.78 & NA\\

AIMD\footnote{From Ref.~\onlinecite{Stich1991}} & 64 & 108.32 &15.5 & 96.60 & 0 & 0.20 &3.20 &NA \\
\end{tabular}
\end{ruledtabular}
\end{table}
\begin{figure*}[t!]
\centering
\includegraphics[width=0.31\textwidth]{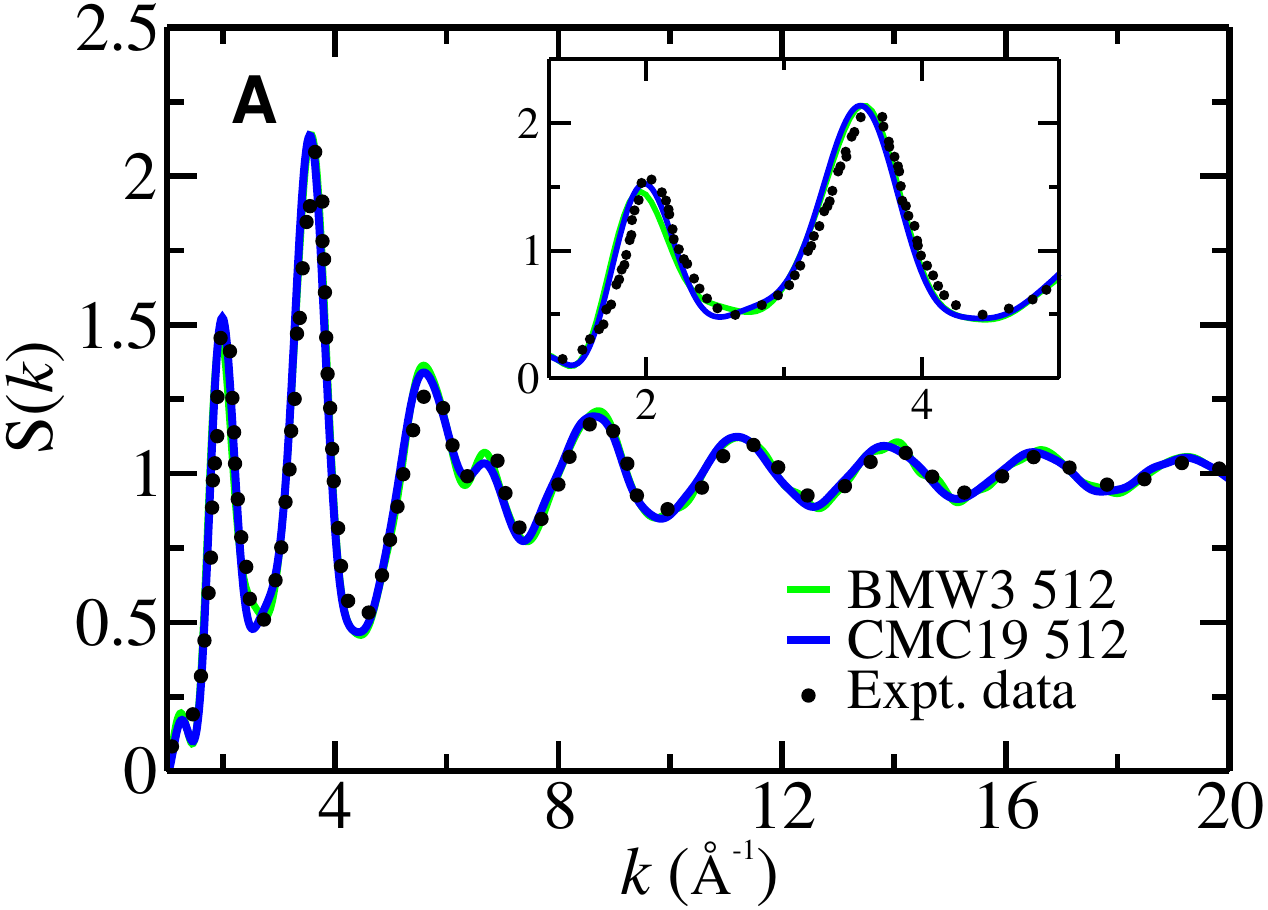} 
\includegraphics[width=0.31\textwidth]{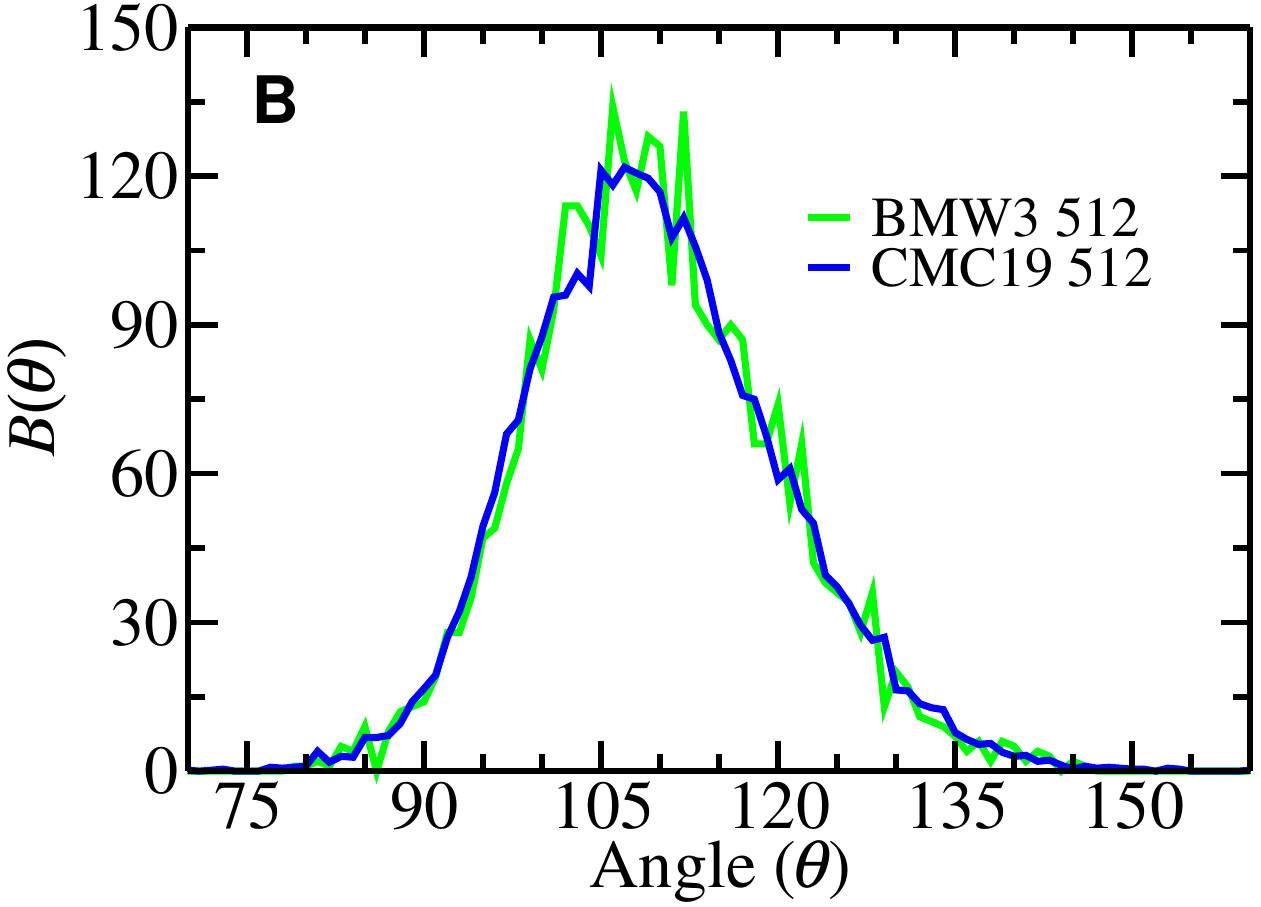} 
\includegraphics[width=0.30\textwidth]{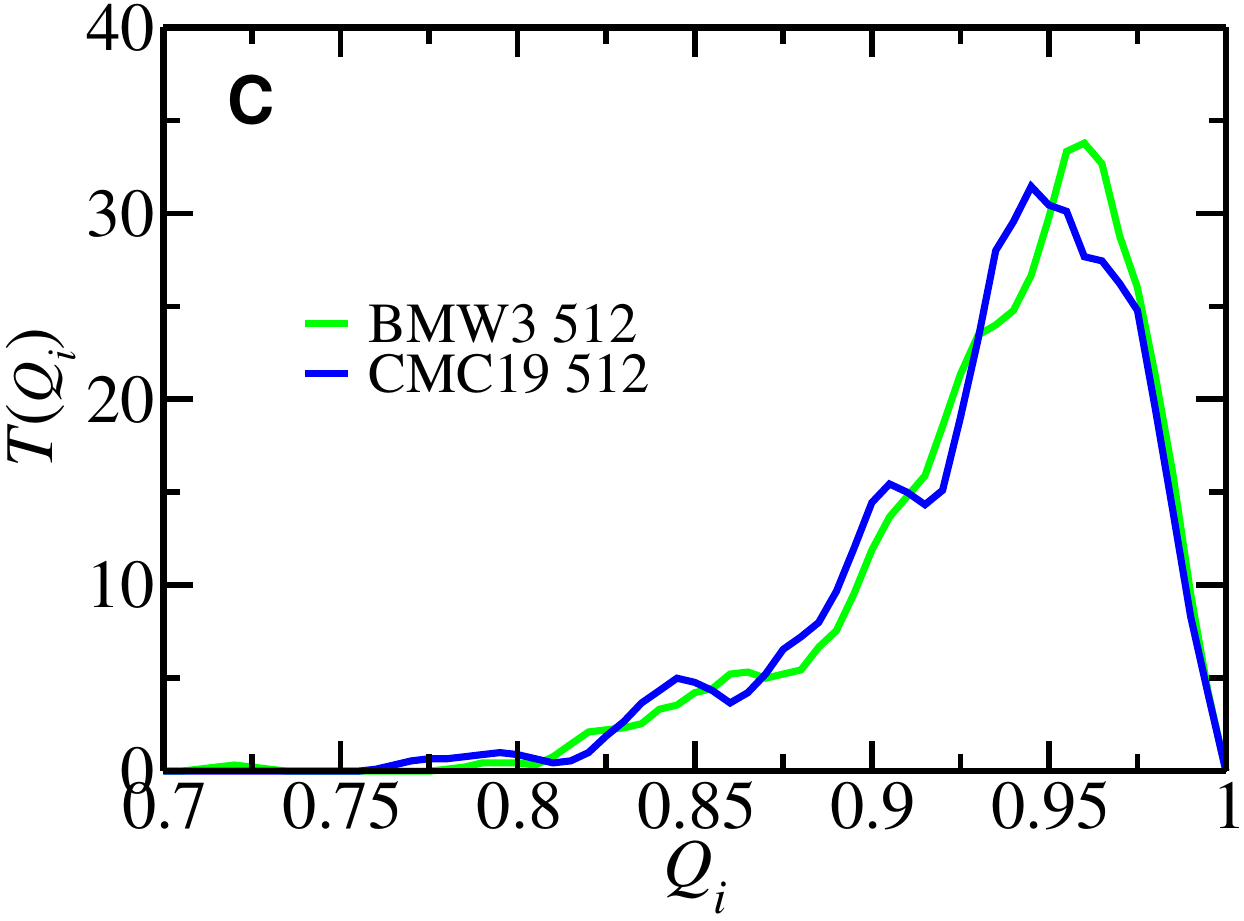} 
\includegraphics[width=0.31\textwidth]{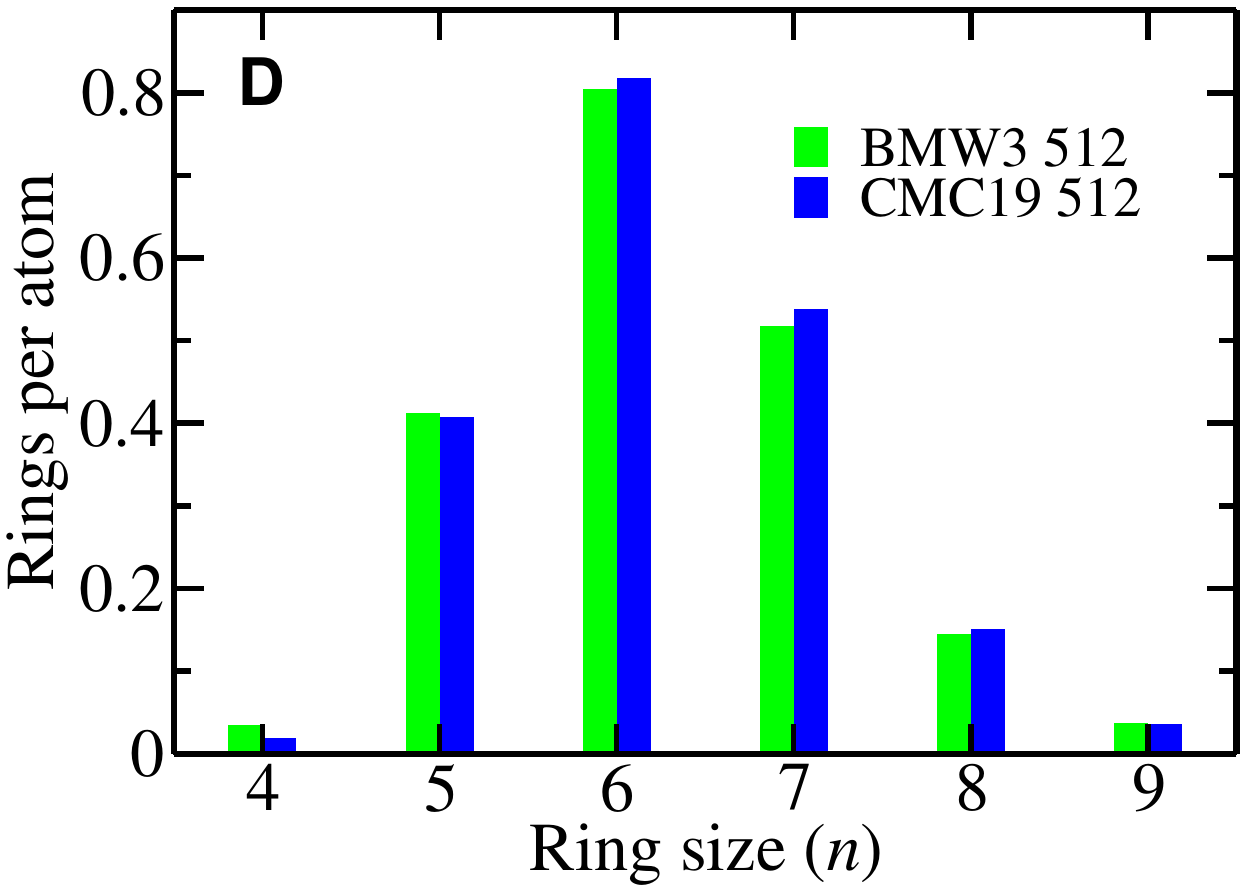} 
\includegraphics[width=0.32\textwidth]{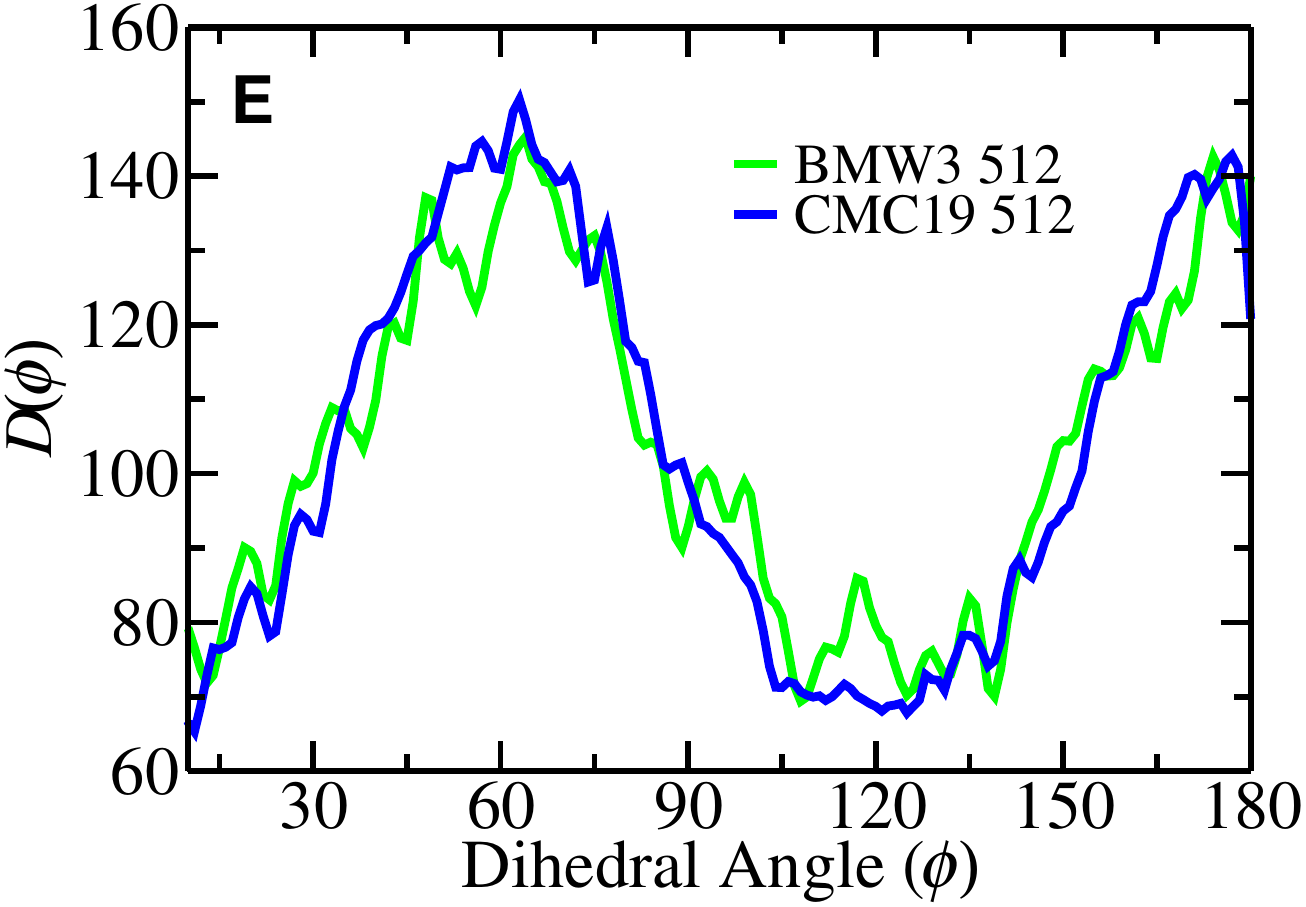} 
\includegraphics[width=0.31\textwidth]{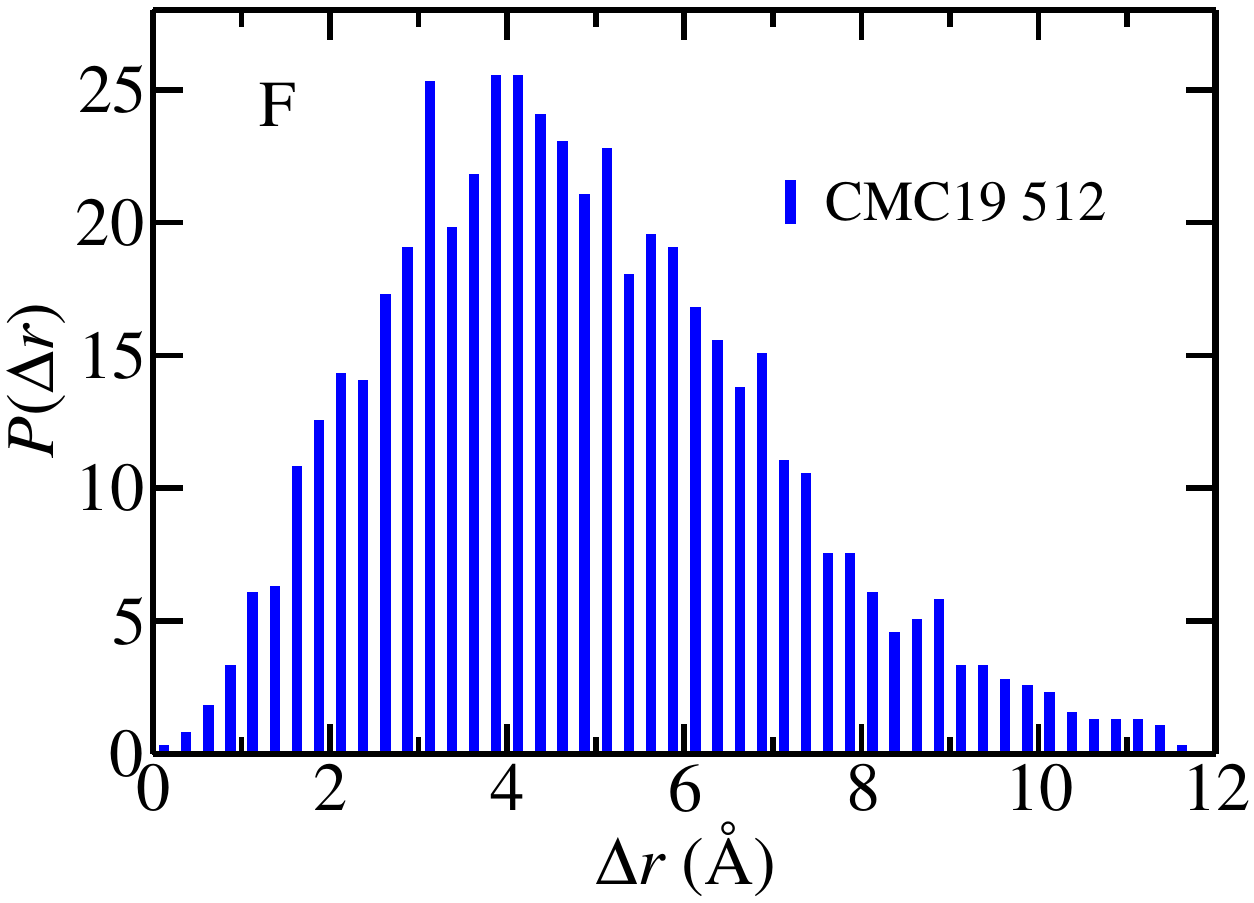} 
\caption{\label{fig1}{\bf Structural properties 
of {\asi} from CMC19 models.} ({\bf A})
The structure factor of a 512-atom CMC19 model (blue) 
and a 512-atom BMW3 model (green). Experimental 
data (black) correspond to as-deposited samples from 
Ref.~\citenum{Laaziri1999}.  An enlarged view 
of the first two peaks is shown in the inset. ({\bf B})
The bond-angle distribution, $B(\theta)$, for a CMC19 
model (blue) and a BMW3 model (green) of identical size.  
({\bf C}) The distribution, $T(Q_i)$, of the local tetrahedral 
order parameter, $Q_i$, for a CMC19 model, along with its 
BMW3 counterpart of identical size. ({\bf D}) The statistics of 
irreducible rings for CMC19 and BMW3 models. ({\bf E}) The 
dihedral-angle distribution, $D(\phi)$, from a CMC19 model
and its BMW3 counterpart, showing the characteristic 
dihedral peaks for tetrahedral ordering near 60{\dg} 
and 180{\dg}. ({\bf F}) The distribution of the resultant 
displacements of atoms during CMC19 simulations for a 
512-atom model.  See text for details. The results 
presented here in {\bf A-F} are all averaged over five (5) 
independent configurations.}
\end{figure*}

Figure \ref{fig1}A presents the structure factor of {\asi} 
obtained from a 512-atom CMC19 model. For comparison, the 
corresponding structure factor from a BMW3 model of identical 
size and from experiments~\cite{Laaziri1999} are also 
included in Fig.\,\ref{fig1}A. 
Owing to the form of the objective function 
in Eq.\,(\ref{chi}), it is not surprising that the structure factor 
from the CMC19 model matches closely with the same quantity from 
the BMW3 model and the experimental data from as-deposited samples in 
Ref.\,\citenum{Laaziri1999}. 
Likewise, the bond-angle distribution in Fig.\,\ref{fig1}B, 
with an RMS bond-angle width of 10.61{\dg}, is also found to 
match very well with its BMW3 counterpart, which is characterized 
by a root-mean-square (RMS) width, $\Delta \theta$, of 10.36{\dg}. 
Assuming that the bond-angle distribution can be approximated 
by a Gaussian function, this value corresponds to a full 
width at half maximum (FWHM) of 23--25{\dg}, which closely 
matches with the measured 
value of the FWHM of about 22-23{\dg}, from Raman 
spectroscopy.~\cite{Beeman1985} It is notable that none of 
the CMC19 models has any 60{\dg} bond angles, and the 
vast majority of the bond angles are narrowly distributed, 
between 90{\dg} and 135{\dg}, as shown in 
Fig.\,\ref{fig1}B. Together with the structure factor, 
the bond-angle distribution and its RMS width play an important 
role in determining the structural quality of tetrahedral 
amorphous semiconducting networks. 

Further characterization of the models is possible by examining the 
local degree of tetrahedrality of the amorphous networks. 
While the average bond angle, $\langle\theta\rangle$, and its 
width, $\Delta \theta$, provide an overall measure of the 
tetrahedral character of a network, the tetrahedral order 
parameter (TOP), $Q_i$, at a site $i$, and its distribution, 
$T(Q_i)$, gives a precise local measure of the tetrahedral 
order. The local TOP can be written as,~\cite{Errington2001}
\be Q_i = 1 - \frac{3}{8}\sum_{\{jik\}} \left(\cos\theta_{jik} + 
\frac{1}{3}\right)^2,  \: \: \langle Q \rangle=\frac{1}{N} \sum_i Q_i, 
\ee 
where the symbol $\{jik\}$ stands for the sum over all possible 
nearest-neighbor angles subtended at the site $i$. It has been 
observed that the electronic quality of a model depends on $T(Q_i)$, 
$\Delta \theta$, and the concentration of coordination 
defects.~\cite{JCP2018} In particular, the local TOP is sensitive 
to the presence of floating bonds (i.e., 5-fold-coordinated atoms), 
and the existence of dangling bonds (i.e., 3-fold-coordinated 
atoms) is readily reflected in the electronic density of states 
near the Fermi level as defect states. 
A comparison of $T(Q_i)$ in Fig.\,\ref{fig1}C clearly 
establishes the similarities between an CMC19 model and 
a BMW3 model, as far as the degree of local tetrahedrality 
is concerned. This observation is consistent with the 
statistics of $n$-fold coordination numbers, $c_n$, given 
in Table \ref{TAB2}.  The marked similarities between the 
CMC19 and BMW3 models, obtained from two distinctly 
different approaches, are indeed remarkable in view of the 
fact that the former do not include any information from 
a total-energy functional.

Having discussed the local structural properties of the 
models, we shall now address to what extent the atomistic 
structures on the medium-range length scale resemble or differ 
from those of the BMW3 models. This is particularly relevant 
for CMC19 models, as the constraint functions in Eq.\,(\ref{chi}) 
do not carry any information beyond the first nearest-neighbor 
distance, and the experimental two-body correlation data, 
$F_{ex}(r;{\mathbf R})$, may not be sufficient to include the 
characteristic structural properties associated with higher-order 
correlation functions. 
We address this by examining the topological connectivity of 
the networks, which involves irreducible rings of various sizes 
and the dihedral-angle distribution. 
Intuitively speaking, given the local tetrahedral character 
of the network, the distribution of $n$-member rings 
($n \ge$ 4) provides some information about the atomistic 
structure on the medium-range length scale,  whereas the 
dihedral-angle distribution should exhibit some characteristic 
features of a (reduced) four-body correlation function.
Figure \ref{fig1}D shows the distribution of irreducible 
rings, of sizes from $n$ = 4 to $n$ = 9, from an CMC19 model 
and a BMW3 model of equal size. 
It is apparent that, despite the absence of 
a total-energy functional, the CMC19 models exhibit
similar topological connectivity as that for the BMW3 
models. 
Likewise, the dihedral angles, involving a chain of 
four consecutive nearest neighbors, are also found 
to be distributed in a similar way for the CMC19 and 
BMW3 models in Fig.\,\ref{fig1}E. 
These results collectively suggest that the two models are 
atomistically similar as far as the topological 
connectivity and the dihedral angles, involving 
the first few neighbors on the length scale of 
3--6 {\AA}, are concerned. 
In this context, it is appropriate to note that, during 
CMC19 simulations, structural formation takes place
via the rearrangement of the great majority of 
atoms (about 92\%) within the first four neighboring 
shells, over a radial distance of up 
to 8 {\AA}. This is evident from Fig.\,\ref{fig1}F, 
where the distribution, $P(\Delta r)$, of the resultant 
atomic displacements, $\Delta r$, is plotted. Here, 
$\Delta r = |\mathbf r_f - \mathbf r_i|$ and 
$\mathbf r_i$ and $\mathbf r_f$ indicate the initial and 
final positions of an atom at the beginning and end of 
simulations, respectively. 
\begin{figure*}[t!]
\includegraphics[width=0.35\textwidth]{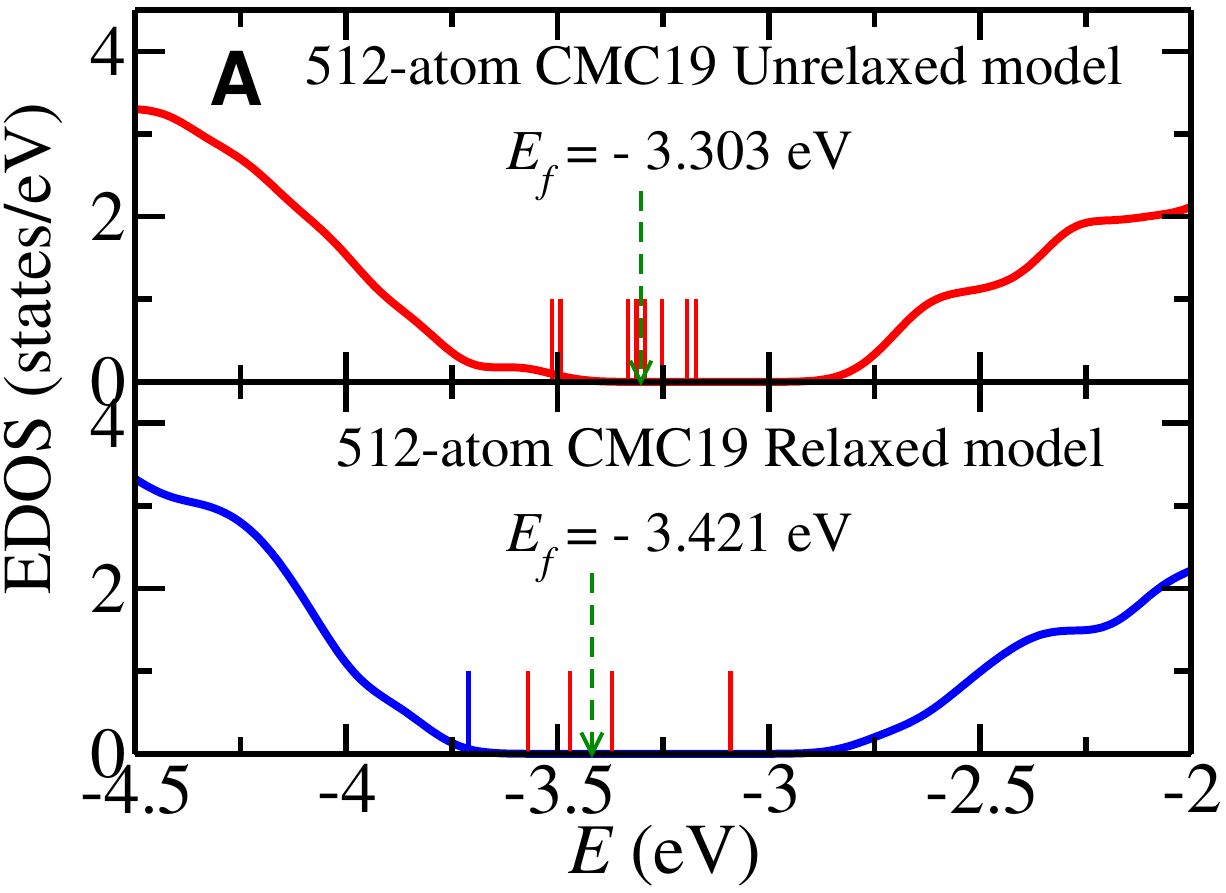}
\includegraphics[width=0.35\textwidth]{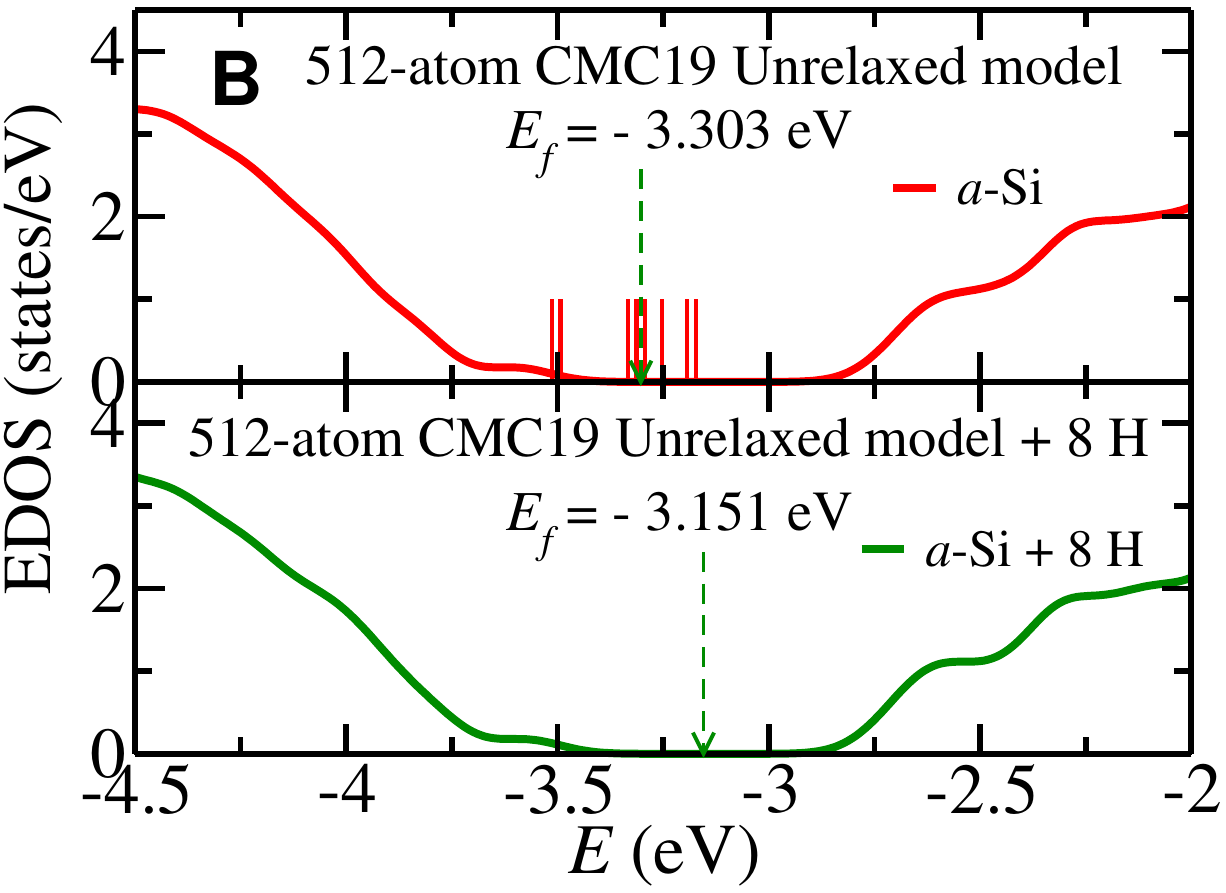}
\includegraphics[width=0.35\textwidth]{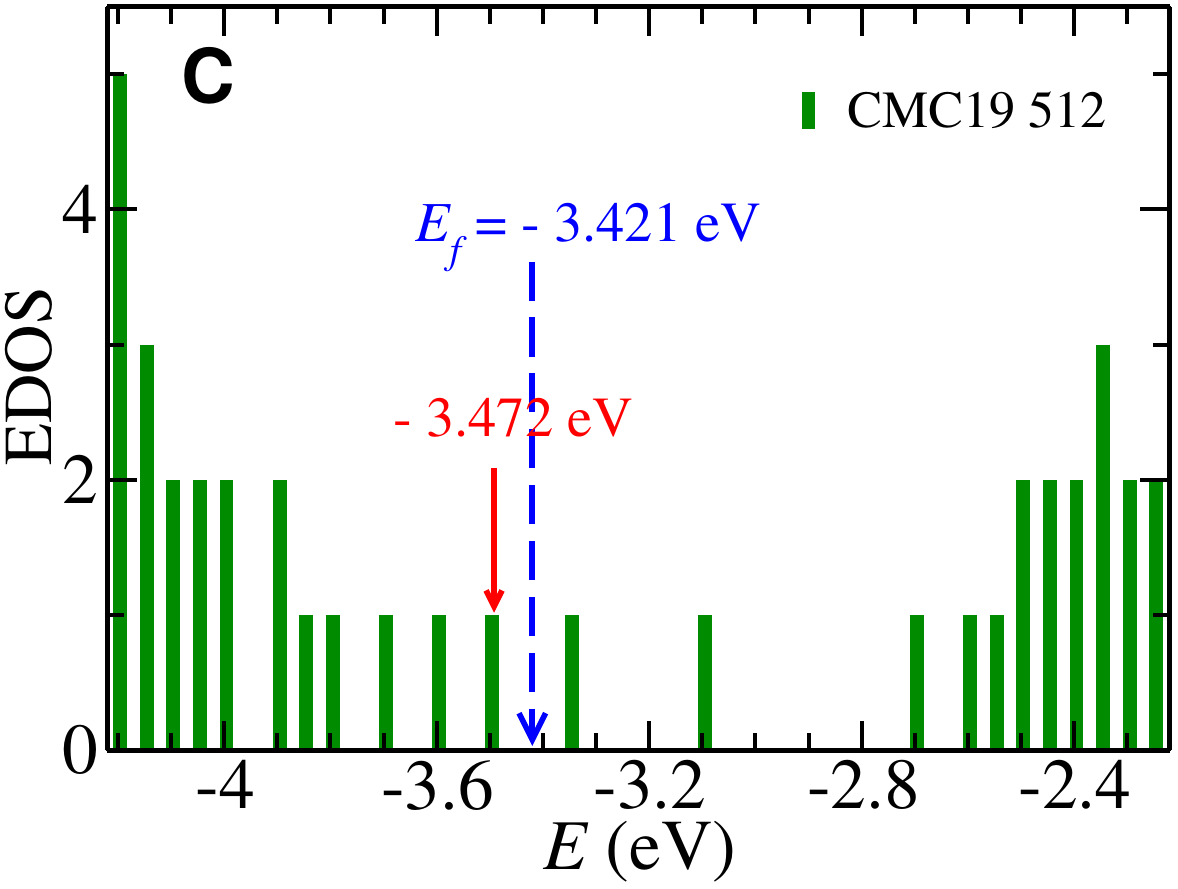}\hspace*{0.5cm}
\includegraphics[width=0.30\textwidth]{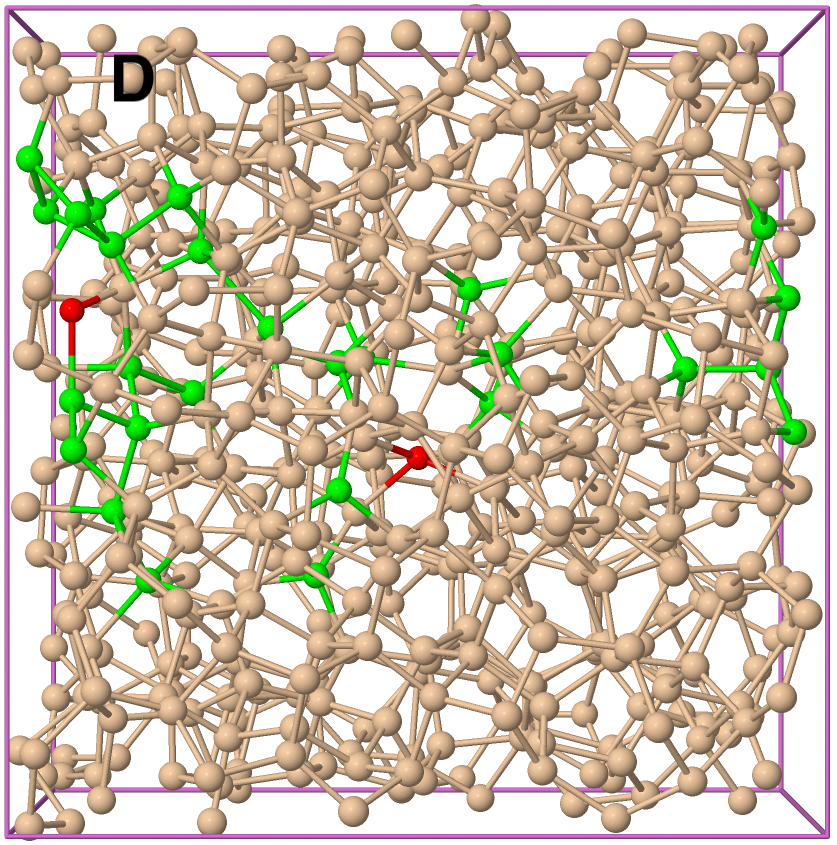}
\caption{\label{fig2}{\bf Electronic density of states (EDOS) of {\asi} 
near the band gap.} ({\bf A}) The EDOS of {\asi} before 
(upper panel) and after (lower panel) 
{\it ab initio} relaxation of the CMC19 model.  A few defect 
states (red) and a band-edge state (blue) are shown as vertical lines in the gap 
region. ({\bf B}) The formation of a clean electronic 
gap in the unrelaxed 512-atom CMC19 model via 
hydrogenation. The upper and lower panels correspond 
to the EDOS before and after H passivation, respectively. 
({\bf C}) A defect state at -3.472 eV and ({\bf D}) the 
associated dangling bonds (i.e., 3-fold-coordinated sites) 
(red) in real space. The other contributing sites, with 4-fold 
coordination, are indicated in green color.
}
\end{figure*} 
In other words, $\Delta r$ indicates the magnitude 
of the resultant displacement of an atom during the entire 
course of an CMC19 run. Figure \ref{fig1}F indicates 
that a considerable number of atoms moved from their 
initial position at the beginning of the simulation 
to the final position at the end of the simulation 
via total atomic displacements as high as 10 {\AA}. 
Since the resultant displacements associated with the 
accepted MC moves are mostly governed by the objective 
function in Eq.\,(\ref{chi}), it is not surprising 
that the radial atomic correlation can 
develop up to a distance of the maximum displacement 
(i.e., $\approx$ 10 {\AA}) of the atoms. This 
observation is indeed reflected in the radial 
distribution function (RDF), where the radial correlations 
between atomic pairs have been found to extend up to a 
distance of 10 {\AA} or more. 

\subsection{Electronic properties of amorphous silicon from CMC19} 

While structural properties of model networks provide 
a wealth of atomistic information,  the most compelling evidence 
of the accuracy and the reliability of a purely 
information-driven method, and the resulting structures 
therefrom, come from its ability to produce 
the correct electronic properties, obtained without 
the use of any total-energy functionals during structural 
formation. For predictive simulations of complex disordered materials 
using experimental data and information only, 
it is of crucial importance to establish the `thermodynamic' 
stability of the models. An effective inverse approach must 
thus be able to produce structural configurations that sit close to a 
stable local minimum of an appropriate total-energy 
functional. However, these primal issues were ignored 
in earlier RMC studies by overemphasizing the importance 
of pair-correlation or structure-factor data, and the 
resulting effects on the three-dimensional structure 
associated with these one-dimensional data. 
Although a few RMC studies~\cite{Biswas2004, Cliffe2010, Cliffe2017,
Pandey2016b} on {\asi}/{\age} did report a narrow bond-angle 
distribution or a high concentration of four-fold atomic 
coordination, those results were obtained at the 
expense of one or the other. Consequently and unsurprisingly, 
none of the models from previous studies could produce a gap 
in the electronic spectrum, either due to a high concentration of coordination 
defects or due to the presence of considerable disorder in 
the bond-angle distribution of atoms, leading to either a 
gapless electronic spectrum or a spectrum with a pseudo-gap.  It 
is therefore necessary to examine the electronic density of 
states (EDOS) of the unrelaxed CMC19 models in order to establish 
that the models are indeed stable prior to structural relaxation 
and that they exhibit an electronic gap in the spectrum 
(see Table \ref{TAB2}). 

Figure \ref{fig2}A show the EDOS of a 512-atom CMC19 model near the 
band-gap region before (upper panel) and after (lower panel) {\it ab initio} 
structural relaxation. Here, we employed the first-principles 
density-functional code {\sc Siesta},~\cite{Siesta2002} using 
double-zeta basis functions and the generalized-gradient 
approximation (GGA).~\cite{PBE} 
\begin{figure*}[t!]
\centering
\includegraphics[width=0.35\textwidth]{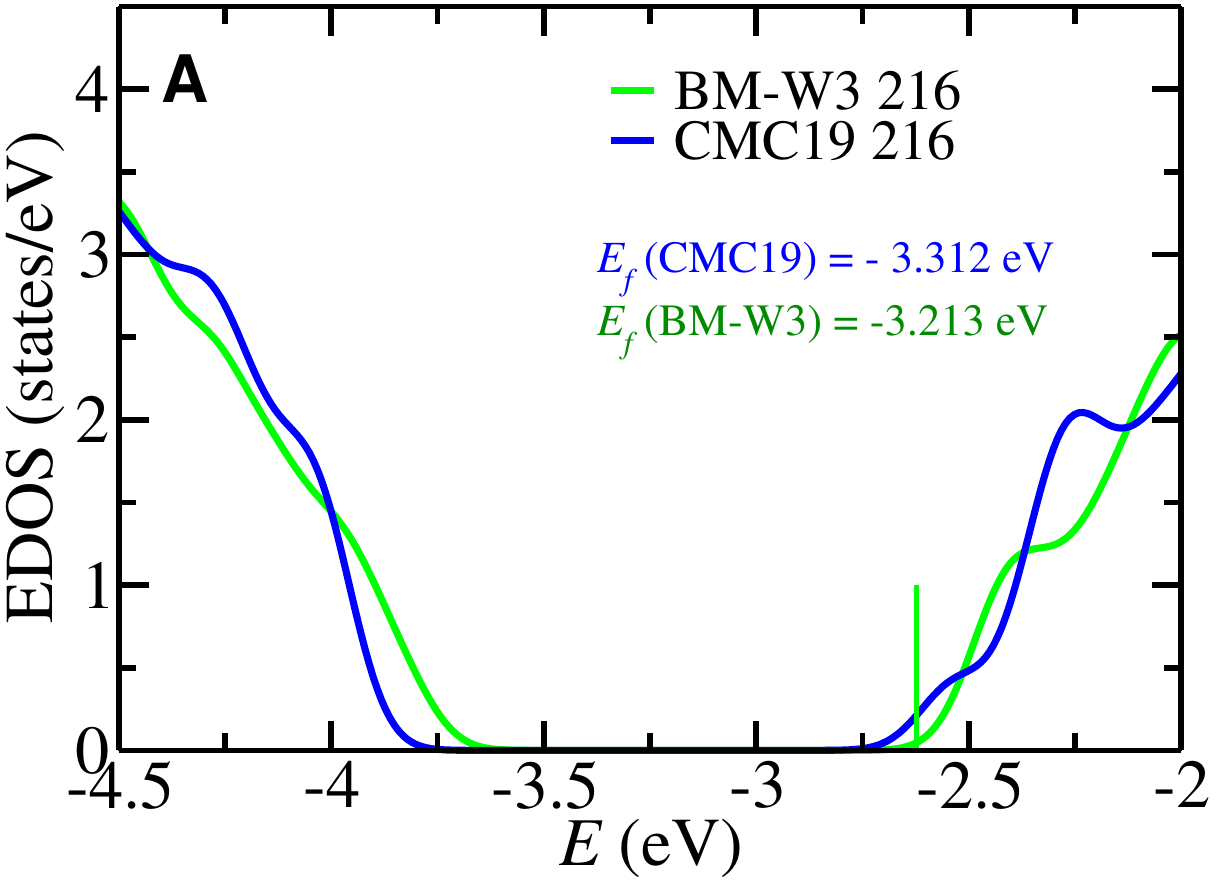}
\includegraphics[width=0.35\textwidth]{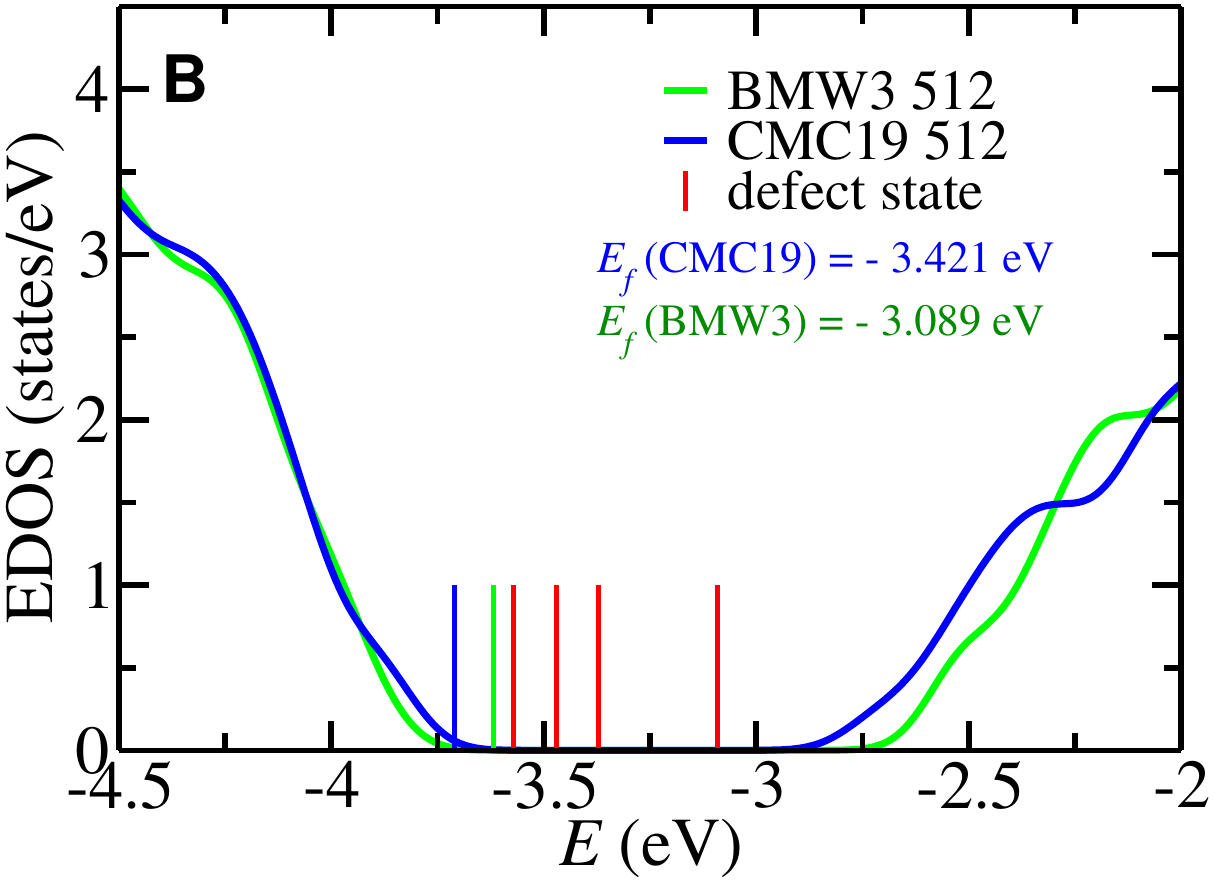}
\includegraphics[width=0.35\textwidth]{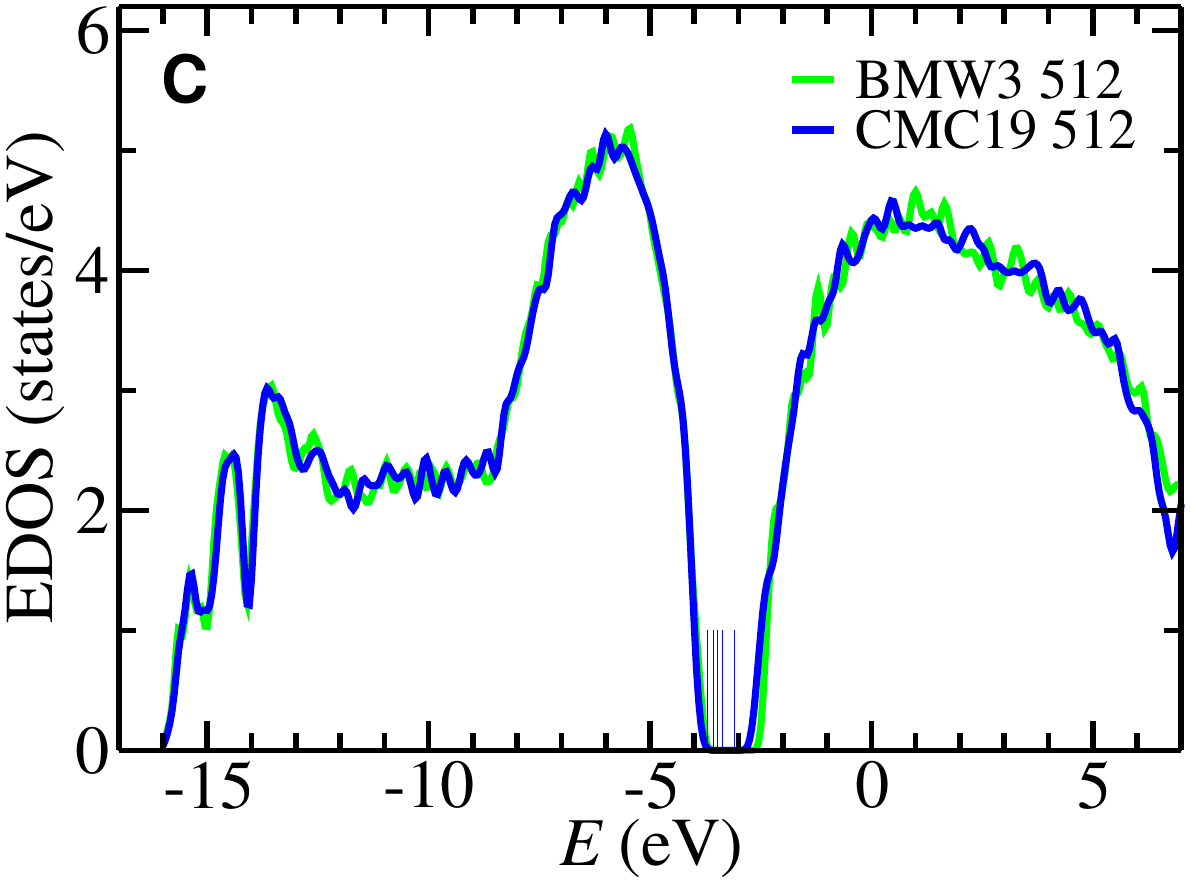}
\includegraphics[width=0.35\textwidth]{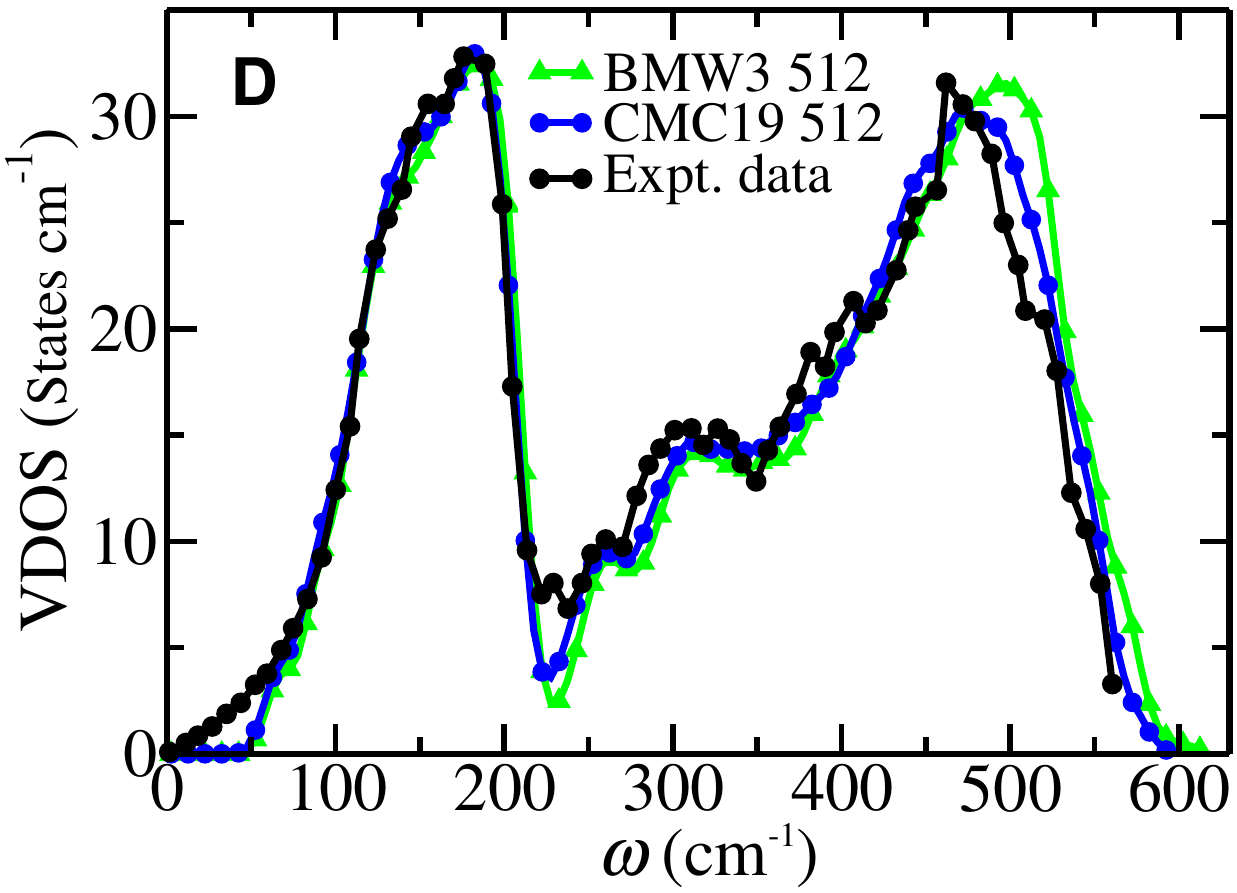}
\caption{\label{fig3}{\bf Electronic and vibrational 
densities of states for {\asi} from CMC19 models.} 
({\bf A}) A 216-atom CMC19 model and its BMW3 counterpart. 
({\bf B}) A 512-atom CMC19 model and BMW3 model, along 
with a few defect states (red) and band-edge states 
(blue and green). ({\bf C}) The full EDOS of {\asi} for a 512-atom 
CMC19 model and a BMW3 model. The vertical lines (blue) 
in the gap region indicate defect states. ({\bf D}) 
The VDOS from a 512-atom CMC19 and a 512-atom BMW3 
model.  Experimental data (black) shown here are from 
inelastic neutron-scattering 
measurements.~\cite{Kamitakahara1984} 
}
\end{figure*}
It is evident that the unrelaxed CMC19 model 
can produce the correct electronic density of states in the 
vicinity of the electronic-gap region. 
Aside from a few defect states, indicated as vertical red lines in 
Figs.\,\ref{fig2}A and \ref{fig2}B, the CMC19 model 
produces a clean spectral gap that 
constitutes a major outcome of this study, previously 
unreported in the literature. A typical defect state 
(at -3.472 eV) near the Fermi level is shown in Fig.\,\ref{fig2}C. 
The state is primarily originated from two dangling 
bonds (DBs) in real space, which are shown in Fig.\,\ref{fig2}D 
in red color. A few neighboring atoms with secondary 
contribution are also indicated in green color. 
A few defect states that appeared in the band-gap 
region can be readily passivated by adding a minute 
amount of hydrogen. This is illustrated in Fig.\,\ref{fig2}B.  
The lower panel of Fig.\,\ref{fig2}B depicts the hydrogen-passivated 
EDOS for the unrelaxed, i.e., {\em static} 512-atom CMC19 model, 
obtained by placing eight (8) H atoms near Si dangling 
bonds.  Hydrogen atoms were so added that the 
local tetrahedral arrangement of the DBs was 
minimally perturbed and the silicon-hydrogen bond 
length was restricted to a distance of 1.55 $\pm$ 0.05 {\AA}. 
The passivation of the Si DBs by H atoms in the static 
512-atom CMC19 model yields a clean electronic 
gap of size 0.85 eV, without {\it ab initio} total-energy 
relaxations, which leads to a high-quality model of 
{\asih} with 1.54 at.\,\% of hydrogen.  This procedure 
can be readily generalized to obtain device-quality 
models of {\asih}, with a varying concentration of 
hydrogen. Thus, the CMC19 approach presented in this 
paper is equally effective in producing models of 
hydrogenated amorphous silicon. 

Since the EDOS near the valence and conduction bands is 
highly susceptible to disorder and the presence of 
coordination defects, particularly dangling bonds 
and the RMS width of bond angles, the presence of 
a clean gap  in the electronic spectrum is often 
considered as the most stringent test of the electronic 
quality of a model. Until recently, with the exception 
of the BMW3 models, there exist no models that can 
exhibit a clean spectral gap in the electronic density 
of states. It is remarkable that the information-driven 
CMC19 models with 216 atoms and 300 atoms exhibit 
a pristine gap, and those with 512 atoms and 
1000 atoms yield a nearly clean gap, in the EDOS 
(see Figs.\,\ref{fig3}A-C and Table \ref{TAB2}), even 
though no total-energy 
functional was employed during the course of minimization 
of the objective function. A comparison of $E_g$ values 
in Table \ref{TAB2}, obtained before and after total-energy 
relaxation of the CMC19 models, firmly establishes that 
the data-driven CMC19 models are energetically stable 
and the latter indeed represent the correct structural 
solution that one strives to obtain from {\it ab initio} 
MD simulations of {\asi}. 

The full EDOS for the 512-atom CMC19 model is presented 
in Fig.\,\ref{fig3}C, along with the results from 
the corresponding BMW3 model.  The EDOS from these 
two models practically matches point-by-point, 
except for a few points near the conduction-band edge, 
as shown in Fig.\,\ref{fig3}C. 
\begin{figure*}[ht]
\includegraphics[width=0.4\textwidth]{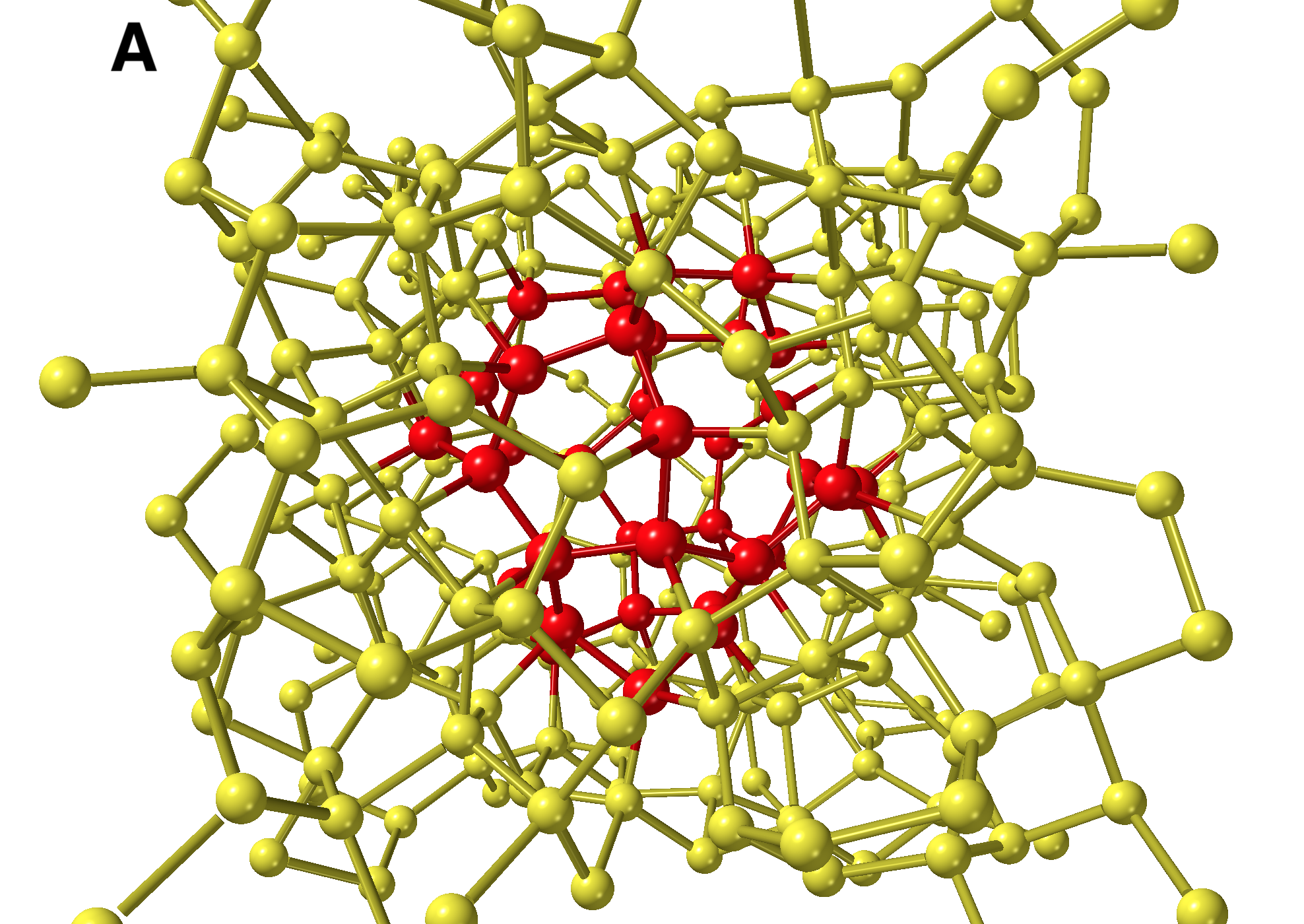}
\includegraphics[width=0.4\textwidth]{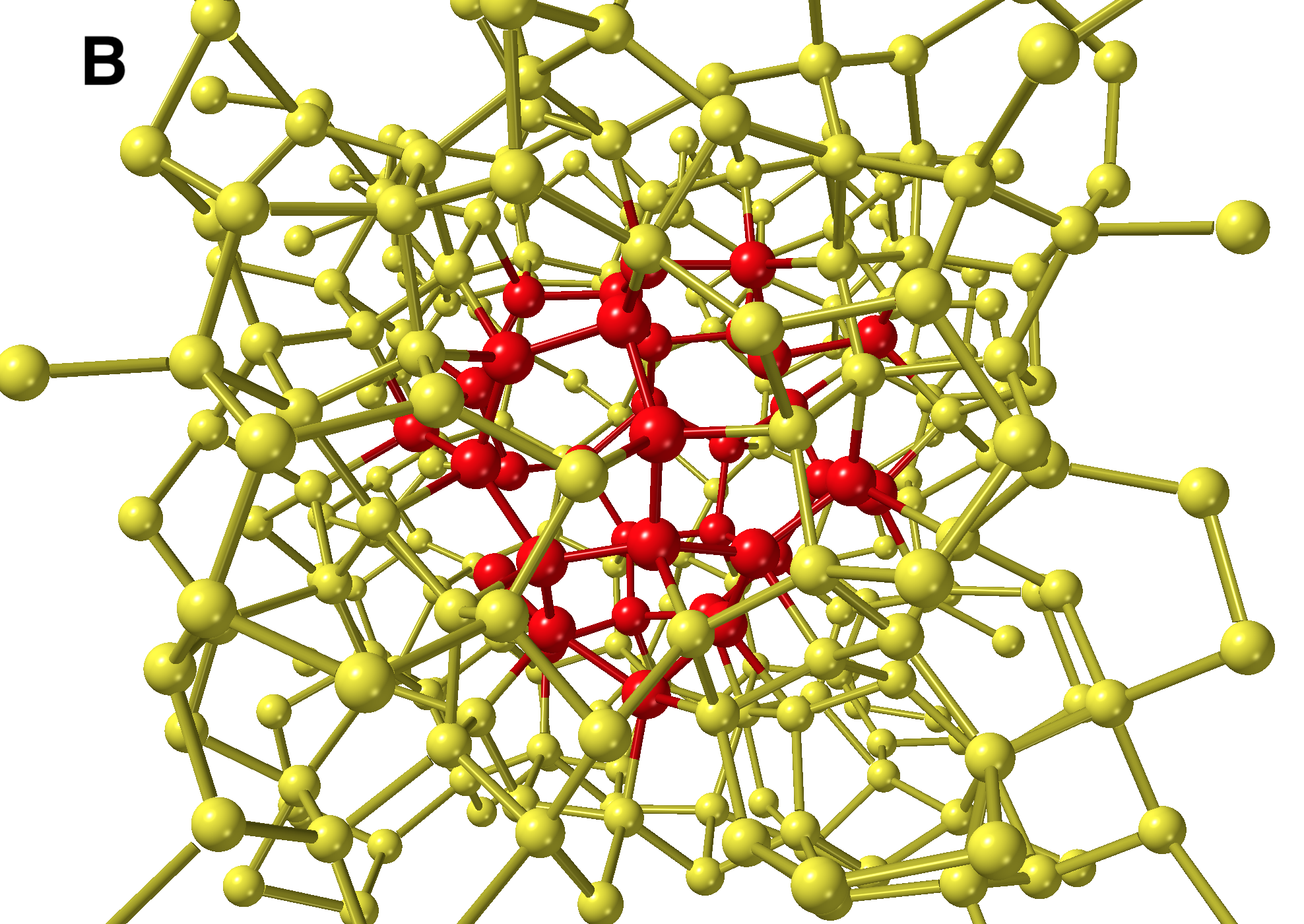} \vskip 0.1cm 
\includegraphics[height=0.3\textwidth, width=0.35\textwidth]{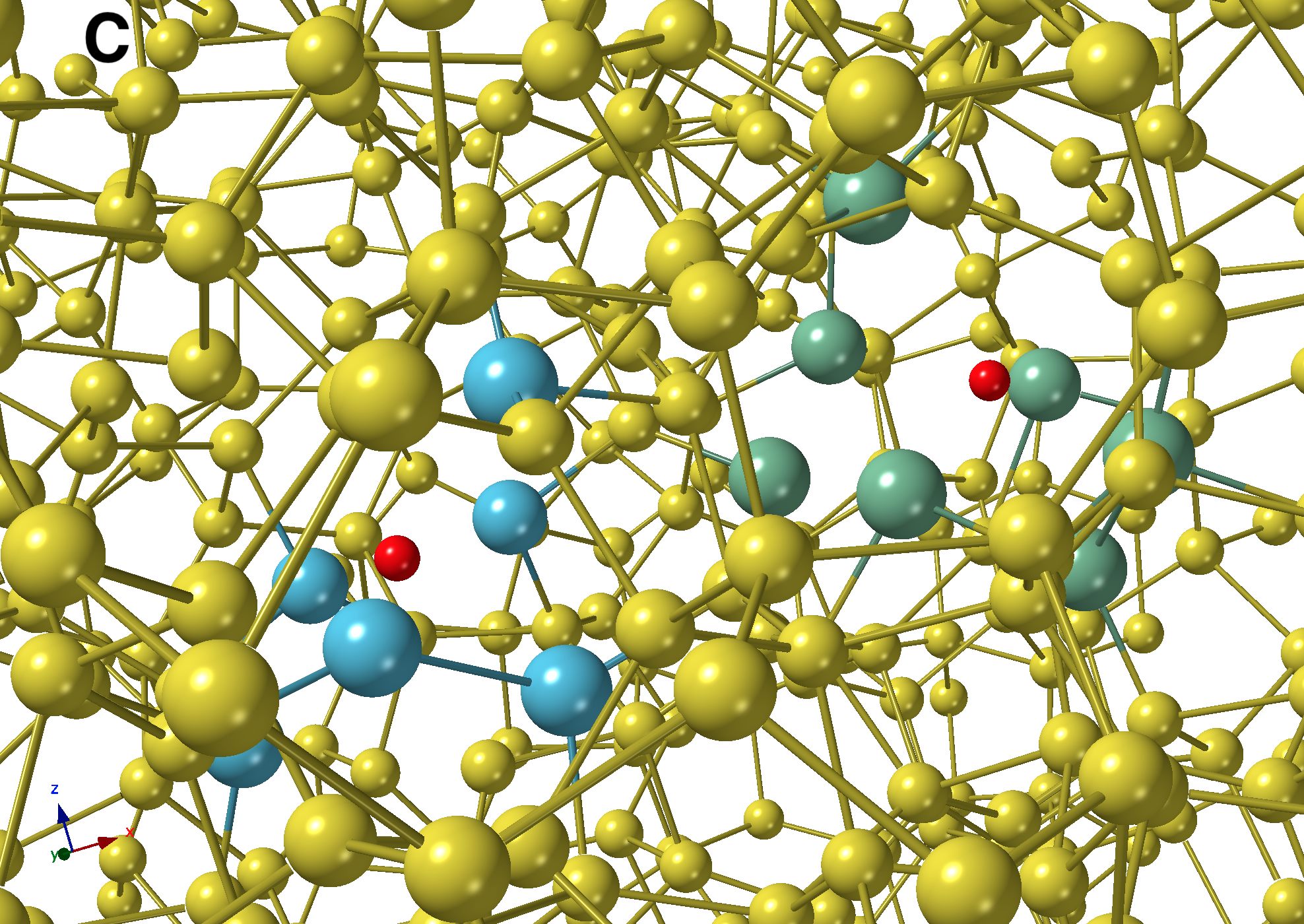}\hskip 0.5cm 
\includegraphics[height=0.3\textwidth, width=0.35\textwidth]{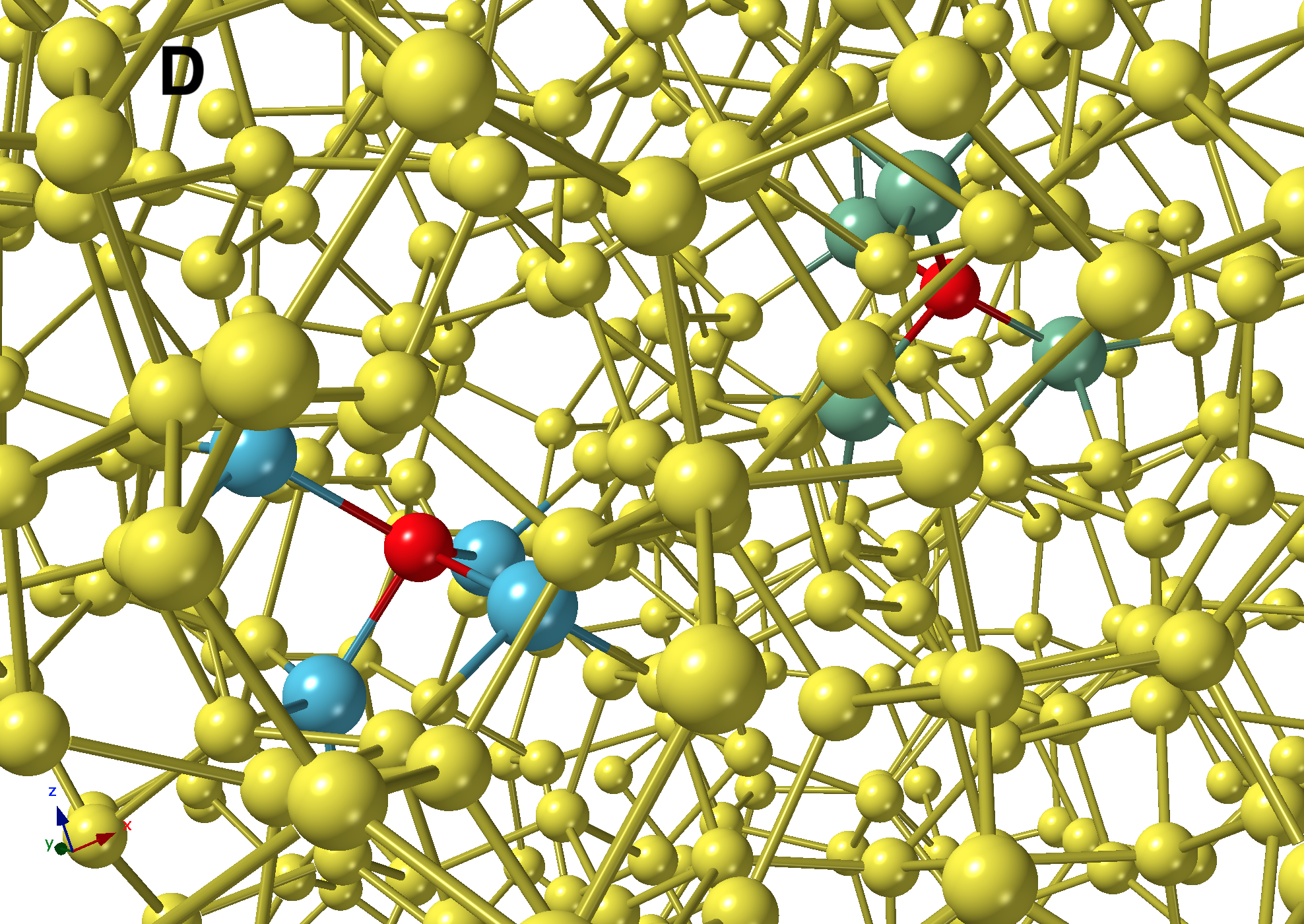}
\caption{\label{fig4} 
{\bf Formation of vacancies and voids in {\asi} via CMC19.} 
({\bf A}) A 300-atom CMC19 model with a void of radius 4 {\AA} at 
the center. The atoms on the void surface (of width 2 {\AA}) 
are shown in red color. ({\bf B}) The same model after {\it ab initio} 
total-energy relaxation, showing the structural stability 
of the void. ({\bf C})  A 512-atom CMC19 model with two monovacancies, 
separated by a distance of 9 {\AA}.  The blue (left) and green 
(right) colors indicate the atoms within the region of 
4 {\AA} from the center of the respective vacancy, which is 
indicated by a small red sphere. ({\bf D}) The {\it ab-initio}-relaxed 
514-atom model obtained by adding a silicon atom (red) at the 
center of each vacancy in Fig.\,\ref{fig4}C. The formation of 
a local 4-fold-coordinated network at/near the vacancy sites 
establishes that the vacancy region corresponds to a 
monovacancy site. The missing Si atoms are shown in red 
color. 
}
\end{figure*}
This minor deviation can be partly attributed to a 
slightly higher value of $\Delta \theta$ for the 
512-atom CMC19 model compared to its BMW3 counterpart. 
The value of the electronic gap, $E_g$, reported 
in Table \ref{TAB2}, was computed by ignoring the 
defect states in the electronic spectrum.  This was 
achieved by examining each of the states in the vicinity 
of the gap region and determining whether it originated 
from coordination defects in real space or not.  
Similarly, the vibrational excitations from a 512-atom 
CMC19 model, which crucially depend on the local 
atomic structure, have been found to agree well with 
those from a BMW3 model of equivalent size and experiments. 
This is evident from Fig.\,\ref{fig3}D, where  the 
vibrational density of states (VDOS) for a 512-atom 
CMC19 model, a 512-atom BMW3 model, and experimental 
data from inelastic neutron-scattering measurements 
from Ref.\citenum{Kamitakahara1984} are presented 
for comparison. The computed values of the VDOS, obtained 
from the harmonic approximation of dynamical matrices, 
match very well with those from experiments and the BMW3 model. 
A small deviation below 50 cm$^{-1}$ can be attributed 
to finite-size effects. 
Assuming a linear dispersion relation, which (generally) 
holds in the limit $k \to 0$, one may expect that the 
normal-mode frequencies below a characteristic frequency 
($\omega_c \propto k_c$) would be absent in finite-size 
models, owing to the presence of a lower wavevector 
cutoff ($k_c \approx \frac{4\pi}{L}$). Additional minor 
deviations from experimental data near 225 cm$^{-1}$ may 
have originated from the application of the harmonic 
approximation in the computation of the VDOS. 

\subsection{Microstructure of realistic samples of {\it a}-Si: 
Modeling vacancies and voids via CMC19}

In the preceding sections, we have discussed the structural, 
electronic and vibrational properties of {\asi}, obtained 
from almost idealized CRN models. As discussed in the 
Method section, the microstructure of realistic 
samples of {\asi} prepared in laboratories can be rather 
complex, depending upon the growth conditions, the method of 
preparation, and the history of the samples, and may not be 
described adequately using the simplistic CRN model of 
{\asi}. However, unlike conventional MD and MC methods, the 
CMC19 approach presented here can effectively address a number 
of microstructural properties of {\asi}, such as voids and 
vacancy defects. We now discuss these properties by 
analyzing {\asi} models,  which are characterized by the 
presence of voids and vacancy-type defects observed in 
laboratory-grown samples of {\asi}. 

The results for 300-atom and 512-atom models using the 
modified objective function in (\ref{chi2}) confirm that 
the approach is very useful to describe microstructural 
properties of {\asi}, starting from a random configuration. Figures 
\ref{fig4}A and \ref{fig4}B show the network structure of a 300-atom 
CMC19 model with a single void of radius of 4 {\AA} 
before and after {\it ab initio} total-energy relaxation, 
respectively. The silicon atoms on the surface of the void, 
between radii 4 {\AA} and 6 {\AA}, are shown in red color 
for visual clarity. 
The structural stability of the void is reflected in 
Fig.\,\ref{fig4}B, which is found to remain intact 
during total-energy relaxations.  
This observation demonstrates that one may include 
voids of varying sizes in the amorphous matrix of {\asi}/{\age} 
so that the void-volume density is consistent with the 
experimentally observed value of 0.02--0.3\%, depending 
upon the method of preparation and conditions.

The vacancy-type defects can also be incorporated in a similar manner 
by producing an array of microvoids of radius 3--4 {\AA}. 
This is illustrated in Figs.\,\ref{fig4}C-D by adding two 
isolated monovacancies in a 512-atom CMC19 model during the 
course of the CMC19 simulations. The vacancy regions are shown 
in Fig.\,\ref{fig4}C by light blue (left) and green 
(right) atoms, which are at distance of up to 4 {\AA} 
from the center of the respective vacancy. The vacancy 
centers are indicated in Fig.\ref{fig4}C by two small 
(hypothetical) red spheres, which are separated by a 
distance of 9 {\AA}. 
The fact that these regions truly represent monovacancies, 
and not microvoids, can be readily verified by placing a 
silicon atom at the center of each vacancy and relaxing 
the resulting (512+2)-atom model using {\it ab initio} 
total-energy functionals. 
It is evident from Fig.\,\ref{fig4}D that the addition of 
two Si atoms passivated the pair of vacancies by restructuring 
the local regions at or near the vacancy sites, which led to 
the formation of a defect-free four-fold-coordinated local 
network. The atoms within a radial distance of 3.2 {\AA} 
from the vacancy centers, are shown as light blue and green 
atoms.  Thus, the pair of isolated vacancies in Fig.\,\ref{fig4}C 
can be viewed as originating from the missing two Si 
atoms, shown in Fig.\,\ref{fig4}D in red color, which are 
separated by a distance of 9.17 {\AA} in the {\it ab initio}-relaxed 
model. It may be noted that the presence of such 1--2\% vacancy-type 
defects in annealed samples of {\asi} was suggested in 
Ref.\,\citenum{Laaziri1999} in order to explain the observed 
value of the average coordination number of 3.88 from 
X-ray diffraction.~\cite{note5} 
In summary, using an augmented form of the objective function, 
the data-driven CMC19 methodology not only provides a means 
to generate accurate structural models of {\asi}/{\age} but 
also to include microstructural properties of the materials observed 
in experiments without employing any total-energy functionals 
and forces. 

\section*{Conclusions}
In this paper, we present a purely data-driven constraint 
Monte Carlo (CMC19) approach that can produce accurate 
structural models of tetrahedral amorphous semiconductors 
without employing a total-energy functional but using diffraction 
data and local structural information only. 
By posing the material-structure determination 
as an inferential problem and addressing the 
problem as an optimization program, we have shown that the problem 
can be solved by inverting diffraction data to generate a 
three-dimensional structural solution, using Monte Carlo 
methods. Owing to its dependence on local information, the approach 
can be implemented efficiently using an order-$N$ algorithm for 
the evaluation of the cost function of a system consisting of 
$N$ atoms. 
An examination of the unrelaxed CMC19 models and their 
{\it ab-initio}-relaxed counterparts shows that the former 
sits very close to a stable local minimum of a quantum-mechanical 
total-energy functional, indicating the thermodynamic 
stability of the information-driven CMC19 models. The hallmark 
of this new constraint-driven order-$N$ approach is that it has 
the ability to produce 100\% defect-free model configurations 
of amorphous silicon, as exemplified by a 216-atom model. 
Comparisons of structural, 
topological, electronic, and vibrational properties of the models 
with those from experiments revealed that the CMC19 models are 
structurally, topologically, and electronically accurate. 
The salient features of the models include 
a narrow bond-angle distribution, with an RMS deviation of 
9--11.5{\dg}, and an ultra-low defect concentration below 1\%, 
which enables the models to exhibit a clean electronic gap of 
size 0.8--1.4 eV. To our knowledge, none of the RMC or 
RMC-derived inverse and hybrid models in the literature so 
far, which employ diffraction data and a total-energy functional, 
can produce the aforementioned structural and electronic 
properties as accurately as the CMC19 models.
The information-based CMC19 approach presented here not only 
can produce overall structural and electronic properties but 
also the microstructural properties of realistic samples of {\asi} 
from experiments, such as voids and vacancy-type defects, 
which cannot be addressed directly using currently available 
computational methods. This observation heralds 
the resolution of the long-standing problem of uniqueness in 
the structural determination of tetrahedral amorphous 
semiconductors via inversion of diffraction data, 
in particular {\asi}, without employing a total-energy functional. 
The study demonstrates that information-driven 
inverse approaches not only can enhance existing methodologies 
for modeling disordered materials, but also offer a directional 
step change in materials computation and radically different approaches 
to the structural determination of disordered materials, 
based on an information paradigm.  

\section*{Materials and Methods}
In our multi-objective constraint optimization approach, 
we begin with an objective function, $\chi^2({\mathbf R})$, 
which includes information from experimental diffraction 
data and a set of structural constraints, 
\be
\chi^2({\mathbf R}) =  \sum_{i=1}\left[ \frac{F_{ex}(q_i) - 
F_c(q_i; \mathbf R)}{\sigma(q_i)}\right]^2 + \sum_{l=1}^{l_m} \lambda_l\, C_l(\mathbf R), 
\label{chi}
\ee
\noindent 
where $F_c(q; \mathbf R)$ correspond to simulated diffraction 
data, either in wavevector space ($q = k$) or in real 
space ($q=r$), obtained from a distribution of atoms 
$\mathbf R$, $\sigma(q_i)$ is the error associated with experimental 
data, $F_{ex}(q_i)$, and $C_l$s are a set of $l_m$ constraints, 
providing additional information on the structural properties 
of the solid. The coefficients $\lambda_l$ are weights, 
which determine the relative strength of each constraint in 
Eq.\,(\ref{chi}). To incorporate constraint information, 
we prescribe a convex function, $C_l = (f_l({\mathbf R}) - f^0_l)^2$, 
where $f_l$ represents a structural variable, associated 
with a configuration $\mathbf R$, and $f^0_l$ corresponds 
to the same for a true but unknown solution $\mathbf R_0$.  
Our goal is to determine an accurate structural solution 
$\mathbf R_0^{\prime}$, which is {\em sufficiently} close 
to $\mathbf R_0$, by simultaneously minimizing $C_l(\mathbf R)$ and fitting the 
computed structure-factor or pair-correlation data with their 
experimental counterpart. Two structural configurations, 
${\mathbf R_0}$ and ${\mathbf R_0^{\prime}}$, are considered to 
be sufficiently close when a minimal number of physical
observables, derived from these configurations, are found
to be almost identical to each other and that these observables 
can be employed to define a structure uniquely.  
For amorphous tetrahedral semiconductors, 
the local chemistry suggests, $f_1 = {\langle\theta\rangle}$,
$f_2 = \Delta \theta$, and $f_3 = c_4$, where 
$\langle\theta \rangle$, $\Delta \theta$ and $c_4$ 
represent the average bond angle, the root-mean-square (RMS) 
deviation of $\theta$, and the percentage of four-fold-coordinated 
atoms in $\mathbf R$, respectively.  The parameters $\lambda$ 
play an important role in the optimization of the objective 
function, $\chi^2(\mathbf R)$, which largely determine the evolution 
of an approximate solution, $\mathbf R_0^{\prime}$, via 
important sampling of the objective function in high-dimensional 
space, using Monte Carlo simulations. 
One frequently chooses $\lambda_l$, by trial and error, so 
that both the objective function and its constraint components 
can be simultaneously optimized during the CMC runs. It may 
be noted that several pareto-optimal sets of $\lambda_l$ exist, 
which suffice to generate good structural solutions. 
The intention here is not to find an optimal 
set of $\lambda$ values, which is analogous to developing a 
three-body potential, but to obtain accurate structural 
solutions consistent with given data sets. 
%
%
To obtain a close structural solution, $\mathbf R_0^{\prime}$, 
the inversion procedure was implemented by minimizing Eq.\,(\ref{chi}), 
using simulated annealing techniques. An initial random 
configuration, consisting of $N$ Si atoms in a cubical box of 
length $L$, was generated so that the density of the model 
corresponds to the experimental density, 2.25 g/cm$^3$, of {\asi}, 
and no two atoms were at a distance less than 2 {\AA}. The latter 
was enforced throughout the course of simulations to maintain an 
excluded volume of radius 1 {\AA} surrounding each Si atom.  
By choosing $F_c(q; {\mathbf R})$ as the pair-correlation function, 
the Metropolis algorithm was employed to accept or reject a 
trial move, $\mathbf R_i \to \mathbf R_f$, following the Metropolis 
acceptance probability $P = \min\left[1,\exp(-\beta\Delta \chi^2)\right]$, 
where $\Delta \chi^2 = \chi^2(\mathbf R_f)-\chi^2(\mathbf R_i)$ 
and $\beta$ (= $1/k_BT$) is an optimization parameter 
or the inverse temperature of the system. 
The system was initially equilibrated at a (hypothetical) 
temperature of 310 K for $10^5$ Monte Carlo steps (MCS), 
which was followed by linear cooling of 
the system from 310 K to 10 K, in steps of 25 K, for 
every $10^5$ MCS at each temperature. The procedure was then 
repeated, from 110 K to 1 K, for at least 5 or more cycles, 
until the RMS width of the bond-angle distribution and the concentration 
of coordination defects reduced to about 11{\dg} and 1\%, 
respectively. The simulations were performed by moving one 
atom at a time and the maximum atomic displacement was 
restricted to 0.15--0.2 {\AA} in order to keep the acceptance 
rate as high as possible.  

The advantage of moving of one atom (or a few atoms) at a time is 
that it provides a means to move atoms in the specific regions 
of interest and to develop an order-$N$ optimization algorithm 
for the computation of $\Delta \chi^2$, associated with the 
displacement of one atom 
(or a few atoms).~\cite{EPL2001} The linear scaling 
was achieved by updating an initially generated 
pair-correlation function and the list of nearest 
neighbors of a (few) selected atom(s) as the simulation proceeds. 
This was particularly necessary to address systems larger 
than 500 atoms. A detailed description of the order-$N$ method will be 
presented elsewhere. Depending upon the temperature, the 
size of the system and the magnitude of the maximum possible displacement 
of an atom during MC simulations, the acceptance rate (of the MC moves) 
was found to vary from 25\% to 50\%. After 
several trial runs, we settled for two sets of $\lambda$ values. 
For 1000-atom models, we used $\lambda_1$ = 2/3, 
$\lambda_2$ = 4/3, and $\lambda_3$ = 2/3, 
whereas the corresponding values for 216-atom, 300-atom, and 512-atom 
models are given by 1/3, 2/3, and 1/15, respectively. Given the 
pareto-optimal nature of the problem, it is possible to employ 
a different set of $\lambda$ values in order to incorporate 
constraints with a varying strength. 
To conduct the configurational averaging of physical properties 
and to demonstrate the reproducibility of results from our method, 
ten independent configurations for each system size were 
generated and studied in our work. Since the results from 
simulated annealing techniques may often vary, depending upon 
the cooling protocol used in simulations, we also employed 
a second cooling scheme with an exponential decay of temperature 
to examine its possible effect on the quality of the final 
solutions. The results from the exponential cooling scheme 
were found to be similar to those from the linear cooling scheme. \\

Having provided an ansatz for reconstructing the three-dimensional 
structure of tetrahedral amorphous semiconductors, without using 
a total-energy functional, via the inversion of experimental 
data in the presence of structural constraints, we now address the 
realistic modeling of microstructure of {\asi}/{\age} observed in 
experiments.  The use of a minimal number of constraints leads to 
a natural solution as continuous random networks (CRN). However, it 
is widely acknowledged that a CRN model cannot provide a comprehensive description 
of {\asi}/{\age}, and the microstructure of {\asi}/{\age} from 
experiments may considerably vary from sample to sample, 
depending upon the method of preparation, 
history of the samples, preparation conditions, etc. Thus, 
while a CRN model provides for the most part a correct description 
of some key structural and electronic properties of {\asi}, it 
cannot produce many important characteristics of laboratory-grown
samples, such as the distributions of defects and microvoids, local 
inhomogeneities, and the presence of different topologically-connected 
regions,~\cite{Gibson2010} observed in experiments. These microstructural 
properties are often addressed by introducing ad hoc structural measures or 
changes in CRN models. For example, the results from fluctuation 
electron microscopy (FEM)~\cite{Gibson2010} are often explained either 
by including small grains of paracrystals~\cite{Gibson2010} or 
nanometer-size voids,~\cite{Biswas_FEM} density fluctuations by 
implanting vacancy-type defects,~\cite{Smets2003,Biswas2011,Abelson1998}
and extended inhomogeneities by voids~\cite{Paudel2018} in 
CRN models. Although these ad hoc measures are found to be largely 
successful in describing the microstructural properties 
of {\asi} and {\age}, associated with various methods of preparation and 
experimental conditions, they cannot nevertheless replace the 
need for a systematic data-driven method. While machine-learning 
(ML) approaches appear to be a promising route to address some 
of these issues, their success crucially depends upon 
the availability of suitable training data and the ability 
of the underlying ML model to obtain an optimal set of 
learning parameters, which involves solving an optimization program 
under a different guise. Likewise, the majority of 
conventional MD simulations tend to produce too many 
dangling-bond defects ($\ge$ 5\%), which often render the MD models 
unsuitable without further treatment, as far as the defect density 
and electronic density of states (EDOS) are 
concerned.~\cite{note5} By contrast, the information-based 
approaches can be generalized by directly including 
appropriate microstructural constraints, guided by experimental 
information or data, to develop models with realistic microstructural 
properties as observed in experiments. 
For example, to generate {\asi} models with a 
given void-volume density, or vacancy-type defects, one may 
include an additional term $g({\bf r}, R_v)$ to $\chi^2({\mathbf R})$ 
and write, 
\be
{\chi^{\prime}}^2 ({\mathbf R}) = \chi^2 ({\mathbf R}) + 
\sum_k \gamma_k \vert g({\bf r-r_k}; R_v)\vert^2, 
\label{chi2}
\ee
where $g(\mathbf{r-r_k}; R_v)$ is a function that produces 
a void, or a vacancy, at $\mathbf{r = r_k}$ of 
linear size $R_v$.  Equation (\ref{chi2}) can be viewed 
as a regularization of the original objective function 
in (\ref{chi}), which can be optimized for a suitable value 
of $\gamma$ and $g({\bf r}, R_v)$. The shape and size 
of the voids can be approximately controlled by choosing 
a suitable functional form of $g(\mathbf{r-r_k}; R_v)$. For example, 
a spherical void or vacancy region can be constructed by 
using an inverse quadratic function or a Fermi function, 
characterized by an appropriate shape parameter, which 
determines the boundary region of a void/vacancy from the 
rest of the network. The efficacy of this approach to describe 
the microstructure of {\asi} is illustrated by providing 
two examples involving voids and vacancy-type defects.

\section*{Acknowledgements}
The work was partially supported by the U.S. National 
Science Foundation (NSF) under Grant Nos.\,DMR 1507166 
and DMR 1507118.

\section*{Author contributions}
P.B. planned and designed research; D.K.L. executed research; 
D.K.L, P.B., R.A. and S.R.E. analyzed data; P.B. and S.R.E. wrote 
the paper.  

\section*{References and Notes}
\input{apaper.bbl}

\end{document}

%% file: apaper.bbl
%